\documentclass[11pt,english,twoside]{article}

\usepackage[T1]{fontenc}
\usepackage[latin1]{inputenc}
\usepackage[english]{babel}
\usepackage{lmodern}
\usepackage{a4wide}
\usepackage{amssymb, amsmath}
\usepackage{graphicx}
\usepackage[all]{xy}
\usepackage{hyperref}
\hypersetup{linktocpage}
\usepackage{dsfont}
\usepackage{mathabx}
\usepackage{mathtools}

\newcommand{\beq}{\begin{equation}}
\newcommand{\eeq}{\end{equation}}
\def\bea#1\eea{\begin{align}#1\end{align}}
\newcommand{\nn}{\nonumber}
\newcommand{\w}{\wedge}
\newcommand{\id}{\mathds{1}}
\newcommand{\ov}{\overline}
\renewcommand{\i}{\ensuremath{\textnormal{i}}}

\def\del {\partial}
\def\d {{\rm d}}
\def\mmm {\mathcal{M}}
\def\eee {\mathcal{E}}
\def\hhh {\mathcal{H}}
\def\p {\phi}
\def\NS {N\!S}

\DeclareMathOperator{\re}{Re}
\DeclareMathOperator{\im}{Im}

\makeatletter
\@addtoreset{equation}{section}
\makeatother

\begin{document}

\begin{titlepage}

\begin{center}

\phantom{\rightline{\small AEI-2015-...}}

\vskip 3.5cm

{\fontsize{16.1}{21}\selectfont \noindent\textbf{New supersymmetric vacua on solvmanifolds} }

\vskip 2.6cm

\textbf{David Andriot}

\vskip 0.9cm

\textit{Max-Planck-Institut f\"ur Gravitationsphysik, Albert-Einstein-Institut,\\Am M\"uhlenberg 1, 14467 Potsdam-Golm, Germany}
\vskip 0.1cm
\textit{Institut f\"ur Mathematik, Humboldt-Universit\"at zu Berlin, IRIS-Adlershof,\\Zum Gro\ss en Windkanal 6, 12489 Berlin, Germany}

\vskip 0.2cm

{\small \texttt{david.andriot@aei.mpg.de}}

\end{center}

\vskip 3.5cm

\begin{center}
{\bf Abstract}
\end{center}

\noindent We obtain new supersymmetric flux vacua of type II supergravities on four-dimensional Minkowski times six-dimensional solvmanifolds. The orientifold $O_4$, $O_5$, $O_6$, $O_7$, or $O_8$-planes and $D$-branes are localized. All vacua are in addition not T-dual to a vacuum on the torus. The corresponding solvmanifolds are proven to be Calabi-Yau, with explicit metrics. Other Ricci flat solvmanifolds are shown to be only K\"ahler.

\vfill

\end{titlepage}

\tableofcontents

\newpage

\section{Introduction and summary}

When trying to relate string theory to real world physics, having extra dimensions is a source of richness as well as complexity. We consider in this paper type IIA and IIB ten-dimensional string theories, through their target space low energy effective descriptions given by the corresponding supergravities. The space-time is a warped product of a four-dimensional Minkowski space-time, and of the six extra space dimensions chosen here to form a compact internal manifold $\mmm$. The four-dimensional physics obtained after dimensional reduction over $\mmm$ is highly dependent on this internal geometry as well as its field content. The resulting four-dimensional theory can also be understood as describing effective four-dimensional light fluctuations around a ten-dimensional vacuum, and thus depends on the vacuum expectation values of the ten-dimensional fields. One of the main results of this paper is to obtain new ten-dimensional vacua of type II supergravities that lead to new, and phenomenologically interesting, four-dimensional physics.

A common feature of four-dimensional theories obtained by such compactifications is the presence of massless scalar fields called moduli. Those and their physical effects are unobserved, so a standard strategy consists in giving them a mass, by stabilizing them in a potential. A typical way to generate a potential through dimensional reduction is to have some fields along internal dimensions with non-zero vacuum expectation values: those are called fluxes. The ten-dimensional vacuum of a type II supergravity is then not only given by a geometry, but also by a non-zero NSNS $H$-flux, or RR fluxes $F_p$. Finding explicit vacua with fluxes is technically involved, but gets simplified when requiring on top that the vacuum preserves some supersymmetry (SUSY). In addition, such SUSY vacua typically lead to SUSY four-dimensional theories, which may be of phenomenological interest. To summarize, we are interested in this paper in SUSY Minkowski flux vacua of ten-dimensional type II supergravities.

Such vacua were first obtained on $\mmm$ being a Calabi-Yau \cite{Giddings:2001yu}, in particular a six-dimensional torus $T^6$. Applying T-duality transformations, vacua on different manifolds called twisted tori can be generated \cite{Kachru:2002sk}. These manifolds are also known as solvmanifolds, a subclass being nilmanifolds, and are built from solvable or nilpotent Lie algebras as explained in section \ref{sec:twtor}. Conditions to obtain SUSY Minkowski flux vacua of type II supergravities were formalised using Generalized Complex Geometry \cite{Hitchin:2004ut, Gualtieri:2003dx}, first in \cite{Grana:2004bg, Grana:2005sn}, and we review them in section \ref{sec:susysol} and appendix \ref{ap:condSUSYvac}. Thanks to these mathematical tools, a wide search for new vacua was performed in \cite{Grana:2006kf} on all six-dimensional nilmanifolds and some further solvmanifolds. Since then, a few more vacua were found on solvmanifolds, as summarized in section \ref{sec:soltwistedtori}. Let us stress that over the years, a huge formalism and set of tools have been developed for flux compactifications. For phenomenology, those should be applied or tested in concrete settings, so it remains important to have explicit vacua. To that end, solvmanifolds offer an interesting playground, since they are very explicit and simple to handle. Overall, there are nevertheless not so many explicit SUSY Minkowski flux vacua of type II supergravities on solvmanifolds.

Among these vacua, one should first distinguish whether the sources of RR fluxes, namely orientifolds, or $O$-planes, and $D$-branes, are smeared or localized. At the supergravity level, the positions of these extended objects are not completely determined in the smeared case, on the contrary to the localized one, which is at least conceptually problematic for the $O$-planes that are not dynamical objects. In practice, the warp factor is then constant or with restricted coordinate dependence. Another distinction to make is whether these vacua can be related by T-duality to a vacuum on the torus. Indeed, such T-dual vacua lead to the same four-dimensional physics that is well-known. To get new physics, the vacuum should rather sit on another ``T-duality orbit'' not connected to (geometric) vacua on the torus; it is then called ``truly new''. The set of known SUSY Minkowski flux vacua on solvmanifolds (see section \ref{sec:soltwistedtori}) is then not very satisfying: apart from one exception, all localized vacua are T-dual to one on the torus, and all truly new vacua have smeared sources. The notable exception was briefly given in section 2.3.1 of \cite{Andriot:2010ju} as a side result, without looking at its T-duals. Here, we reproduce with more details this localized vacuum in section \ref{sec:newsolO6}, and find many more, presented in section \ref{sec:newsol}. In section \ref{sec:Td}, we show that all these localized SUSY Minkowski flux vacua are truly new.\\

To reach this result, we first study in details localized vacua on nilmanifolds (particular examples of solvmanifolds) in section \ref{sec:solnil}. In particular, we present in section \ref{sec:sol2} a vacuum with an $O_6$-plane and an orthogonal SU(2) structure: it is the first localized SUSY Minkowski flux vacuum on a solvmanifold with such attributes. It is however very likely to be T-dual to a vacuum on the torus. We further discuss in section \ref{sec:propsources} properties of these vacua, in particular the constant terms present in the fluxes and Bianchi identities. We relate this feature to a property of the internal subspace wrapped by the sources: the smeared volume form of this subspace $\widetilde{{\rm vol}}_{||}$ is not closed \eqref{wrappedsubspace}. This property surprisingly holds for many vacua, for instance all those of \cite{Grana:2006kf}. This is however not a generic property, but is rather due to the method used to find vacua. We then discuss vacua for which $\d \widetilde{{\rm vol}}_{||} = 0$: up to few assumptions, their fluxes would not contain constant terms but only be given by derivatives of the warp factor, while the internal manifold would be Ricci flat. This is in a sense realised by the $O_3$ vacua of \cite{Giddings:2001yu} on the torus, and we are interested in finding others on different solvmanifolds.

Ricci flat solvmanifolds are discussed in section \ref{sec:flattwtor}. As shown in \cite{Grana:2013ila}, only three algebras lead to such manifolds: they correspond to $s_1$, $s_2$ and $s_3$ introduced in \eqref{s1s2s3}. These algebras have special properties: they are in one-to-one with their solvable group, but the solvmanifold obtained after further dividing by the lattice differs according to the lattice chosen. Some choices in particular lead to a torus, but other lattices fortunately give different solvmanifolds, as we detail. This is not problematic for the search of the vacua of interest that we perform on these manifolds. We do not find any with SU(3) or orthogonal SU(2) structure on solvmanifolds corresponding to $s_1$ and $s_2$, but we obtain several on those related to $s_3$. We summarize these new localized SUSY Minkowski flux vacua in \eqref{figuresols3}, indicating the $O$-plane and its directions as well as the type of structure. They are gathered in two families: vacua in one family should be T-dual to each other; the T-dualities and their directions are indicated by the arrows.
\beq
\xymatrix{ \mbox{SU(2)}_{\bot}\ O_8~//~ 12345 \ar@{<->}[d]^{34} &  & \\
\mbox{SU(3)}\ O_6~//~ 125 \ar@{<->}[d]^{12} &  & \mbox{SU(2)}_{\bot}\ O_6~//~ 125 \ar@{<->}[d]^{6} \\
\mbox{SU(2)}_{\bot}\ O_4~//~ 5 \ar@{<->}[d]^{6} &  & \mbox{SU(3)}\ O_7~//~ 1256 \\
\mbox{SU(3)}\ O_5~//~ 56 &  & } \label{figuresols3}
\eeq
These vacua are presented in section \ref{sec:newsol}, and we believe that more can be found: as explained there, we do not aim at an exhaustive search, but rather obtain one example for each $O$-plane and discuss its specificities. We show in section \ref{sec:Td} and appendix \ref{ap:Td} that these vacua are not T-dual to geometric vacua on the torus, and are thus truly new.

Last but not least, we study in section \ref{sec:CY} the geometric properties of solvmanifolds corresponding to $s_1$, $s_2$ and $s_3$, all of them being not simply connected. For $s_3$, the particularity of our new vacua makes it possible to take a smeared and fluxless limit that preserves the underlying geometry. In this limit, the SUSY conditions for an SU(3) structure with an $O_5$, $O_6$, or $O_7$, become simply those characterising a Calabi-Yau. This way, and after discussion, we prove that all solvmanifolds corresponding to $s_3$ are Calabi-Yau. Interestingly, this provides examples (including the torus) of Calabi-Yau with explicit metric, even though we explain that their holonomy group is trivial. We argue that these are the only solvmanifolds being Calabi-Yau. We show that those corresponding to $s_1$ and $s_2$ are only Ricci flat K\"ahler manifolds, and study their SU(3) structure torsion class $W_5$.

\section{Background material}

\subsection{Conditions for supersymmetric Minkowski flux vacua}\label{sec:susysol}

We are interested in finding Minkowski vacua of ten-dimensional type II supergravities, with non-trivial fluxes, preserving (at least) ${\cal N}=1$ SUSY in four dimensions. The conditions for such vacua were formulated in \cite{Grana:2004bg, Grana:2005sn} using Generalized Complex Geometry (GCG) \cite{Hitchin:2004ut, Gualtieri:2003dx}, while a first set of new solutions were obtained thanks to this formulation in \cite{Grana:2006kf}. A review on the physics use of this mathematical framework can be found in \cite{Koerber:2010bx}. Here, we briefly present the formalism needed, allowing us to fix notations; we follow conventions of \cite{Grana:2006kf, Andriot:2010sya}.

The ten-dimensional space-time is taken to be a warped product of an external four-dimensional maximally symmetric space-time (along directions $\d x^{\mu}$) and an internal six-dimensional compact manifold $\mmm$ (along directions $\d y^m$). The metric is written accordingly
\beq
\d s^2= e^{2A(y)} g_{\mu\nu} (x) \d x^\mu \d x^\nu + g_{mn} (y) \d y^m \d y^n \ ,
\eeq
where we call for simplicity $e^A$ the warp factor, and $g_{\mu\nu}$ will eventually be the Minkowski metric. The other fields of ten-dimensional type II supergravities, formulated with the democratic formalism, get as well a compactification ansatz. The dilaton $\p$ depends only on internal coordinates. The NSNS $3$-form flux $H$ is taken to have purely internal components, while the non-trivial part of the RR fluxes\footnote{The ten-dimensional self-dual RR fluxes of the democratic formalism, gathered in a polyform $F^{10}$ analogous to $F$ of \eqref{RRIIA} and \eqref{RRIIB}, are related to those internal $F$ by $F^{10}= F \pm {\rm vol}_4 \w \lambda (* F)$, where ${\rm vol}_4 $ is the warped four-dimensional volume form, and the signs and $*$ are defined as for \eqref{SUSY}.} is captured by $p$-forms $F_p$ on $\mmm$, gathered as
\bea
\textrm{IIA}&:\ F=F_0+F_2+F_4+F_6\ ,\ \lambda(F)=F_0-F_2+F_4-F_6 \ , \label{RRIIA} \\
\textrm{IIB}&:\ F=F_1+F_3+F_5\ ,\ \lambda(F)=F_1-F_3+F_5 \ . \label{RRIIB}
\eea
Vacua preserving SUSY should have vanishing fermionic (gravitino, dilatino) SUSY variations. This gives non-trivial differential conditions on the ten-dimensional fermionic SUSY parameters $\epsilon^{1,2}$: the Killing spinor equations. These conditions are then split according to the compactification; in particular, one takes for an ${\cal N}=1$ vacuum
\beq
\mbox{IIA}\ \begin{cases} \epsilon^1 = \zeta_+ \otimes \eta^{1}_+ + \zeta_- \otimes \eta^{1}_- \\ \epsilon^2 = \zeta_+ \otimes \eta^{2}_- + \zeta_- \otimes \eta^{2}_+ \end{cases} \ ,\qquad \mbox{IIB}\ \begin{cases} \epsilon^1 = \zeta_+ \otimes \eta^{1}_+ + \zeta_- \otimes \eta^{1}_- \\ \epsilon^2 = \zeta_+ \otimes \eta^{2}_+ + \zeta_- \otimes \eta^{2}_- \end{cases} \ ,
\eeq
where $\zeta$ is the external SUSY parameter, $\eta^{1,2}$ are internal spinors, and $\pm$ denote the chiralities. Although not strictly necessary, the internal spinors are considered globally defined on $\mmm$. This provides topological conditions on the internal manifold: if the two spinors are the same, the structure group of the tangent bundle is reduced to SU(3); if they differ, it is further reduced to SU(2). More generally, the structure group of the generalized tangent bundle of GCG is SU(3)$\times$SU(3). After the split, the non-trivial parts of the differential conditions to preserve ${\cal N}=1$ SUSY are internal, and can be reformulated in terms of
\beq
\Phi_{\pm} = \eta^{1}_+ \otimes \eta^{2\dag}_{\pm} \ ,\ \begin{cases} \mbox{IIA}: \Phi_1=\Phi_+\ ,\  \Phi_2=\Phi_- \\ \mbox{IIB}: \Phi_1=\Phi_-\ ,\  \Phi_2=\Phi_+ \end{cases} \ . \label{purespinordef}
\eeq
These bispinors are also (pure) spinors on the generalized tangent bundle of GCG. In addition, they can be understood as polyforms on $\mmm$ thanks to the Fierz identity and the Clifford map. Their corresponding expression depends on the structure group. Locally or for constant structures, one should distinguish three cases: an SU(3), an orthogonal (or static) SU(2), or an intermediate SU(2) structure. The corresponding polyforms are respectively
\bea
{\mbox SU(3)}:& \ \Phi_+ = \frac{e^A}{8} e^{\i \theta_+} e^{-\i J} \ ,\ \Phi_- = - \frac{e^A}{8} \i e^{\i \theta_-} \Omega \ , \label{purespinorstruct}\\
{\mbox SU(2)}_{\bot}:& \ \Phi_+ = -\frac{e^A}{8} \i e^{\i \theta_+} \omega\w e^{\frac{1}{2} z\w \ov{z}} \ ,\ \Phi_- = -\frac{e^A}{8} e^{\i \theta_-} z\w e^{-\i j} \ ,\nn\\
{\mbox SU(2)}_{\angle}:& \ \Phi_+ = \frac{e^A}{8} e^{\i \theta_+} k_{||} e^{\frac{1}{2} z\w \ov{z} - \i j -\i \frac{k_{\bot}}{k_{||}} \omega} \ ,\ \Phi_- = - \frac{e^A}{8} e^{\i \theta_-} k_{\bot} z\w e^{- \i j +\i \frac{k_{||}}{k_{\bot}} \omega} \ ,\nn
\eea
where $k_{||},\ k_{\bot}$ are constant non-zero coefficients, $\theta_{\pm}$ are constant phases, $j$, $J$ are (1,1)-forms and $z$, $\omega$, $\Omega$ are (1,0)-, (2,0)-, and (3,0)-forms, with respect to an almost complex structure $I$. These forms can be expressed in terms of the internal spinors. Here the norm of the spinors has been fixed with the warp factor, as required in presence of an orientifold. The SUSY differential conditions, for an external Minkowski space-time, are then formulated as \cite{Grana:2005sn, Grana:2006kf}
\bea
& (\d -H\w)(e^{2A-\phi}\Phi_1)=0 \ , \label{SUSY}\\
& (\d -H\w)(e^{A-\phi}\re(\Phi_2))=0 \ , \nn\\
& (\d -H\w)(e^{3A-\phi}\im(\Phi_2))=\pm\frac{e^{4A}}{8}  *\lambda(F) \ , \nn
\eea
where the last sign is $+$ for IIA and $-$ for IIB and $*$ is the Hodge star for the internal metric. To guarantee in addition the topological requirement, namely that $\Phi_{\pm}$ provide an SU(3)$\times$SU(3) structure, they need to verify compatibility conditions.

The fluxes have on top to satisfy their Bianchi identities (BI). Here we consider no $\NS5$-brane neither Kaluza-Klein monopole, while RR sources are space-time filling $D$-branes $D_p$ and orientifold planes $O_p$. We do not consider world volume flux ${\cal F}$ on the $D$-branes. Finding Minkowski vacua with fluxes on a compact manifold requires negative contributions to the BI \cite{Maldacena:2000mw, Giddings:2001yu}, given here for RR by $O$-planes. The BI are then given by\footnote{The BI for the geometric flux \cite{Villadoro:2007yq, Andriot:2014uda} in absence of a Kaluza-Klein monopole will be automatically satisfied here, as we will work on manifolds based on Lie algebras.}
\beq
\d H=0 \ ,\quad (\d- H\w) F=  - \sum_{O_p} 2^{p-5}\ c_p\ {\rm vol}_{\bot} \delta(y_{\bot}) + \sum_{D_p} c_p\ {\rm vol}_{\bot} \delta(y_{\bot}) \ , \label{BI}
\eeq
where ${\rm vol}_{\bot} \delta(y_{\bot})$ is a localized transverse volume form to a source, and $c_p$ is a positive normalisation factor including the $D$-brane tension. We consider RR sources compatible with the bulk SUSY, meaning preserving part of it. Provided the (bulk) background is SUSY, such sources are calibrated, i.e. their energy is minimized \cite{Koerber:2005qi, Martucci:2005ht, Koerber:2006hh, Koerber:2007hd} (see also \cite{Witt, Martucci:2011dn}). Their (world volume) volume form ${\rm vol}_{||}$ is then given by the pullback of $\im(\Phi_2)$
\beq
P[ \im(\Phi_2) ] = \frac{e^A}{8} {\rm vol}_{||} \ .\label{calib}
\eeq
We will check this calibration condition. Finally, the orientifold imposes projection constraints to be verified: $\Phi_{\pm}$ or equivalently the structure forms should transform in a specific manner with respect to its involution $\sigma$, and the manifold should be compatible with $\sigma$, which boils down here to $\sigma$ preserving the underlying algebra.

For a field configuration to be a vacuum, it should a priori solve the equations of motion. Given the present compactification ansatz, it has been shown that satisfying the SUSY conditions and the BI is however sufficient to get an ${\cal N}=1$ SUSY Minkowski vacuum \cite{Lust:2004ig, Gauntlett:2005ww, Grana:2006kf, Koerber:2007hd}. So we conclude by summarizing the conditions to be verified in practice to get such a vacuum:\\
- compatibility of the manifold with the orientifold projection;\\
- compatibility of the structure forms with the orientifold projection;\\
- compatibility conditions of the structure group;\\
- SUSY conditions \eqref{SUSY};\\
- Bianchi identities \eqref{BI};\\
- calibration of the sources \eqref{calib}.\\
Some pragmatic details on these conditions are provided in appendix \ref{ap:condSUSYvac}. More explanations on this method can also be found in \cite{Andriot:2010sya}.

\subsection{Basics and notations for solvmanifolds}\label{sec:twtor}

In this paper, we are interested in internal manifolds $\mmm$ being solvmanifolds, commonly named twisted tori in physics. These are compact manifolds obtained by dividing a solvable Lie group by a lattice (for a review, see \cite{Andriot:2010ju}). A subclass of those are nilmanifolds, obtained similarly from nilpotent Lie groups. The resulting manifolds are twisted tori: they are made of fibrations of tori, the only one with trivial fibration being the torus itself. We give here some basic elements on these manifolds and fix some related notations needed later on. An account on vacua found with such an $\mmm$ is provided in section \ref{sec:soltwistedtori}.

Consider a Lie algebra of vector basis $\lbrace E_a \rbrace$ and structure constants $f^c{}_{ab}$, defined by
\beq
\lbrack E_b , E_c \rbrack = f^a{}_{bc} E_a \ . \label{algebra}
\eeq
Solvmanifolds are built starting with a solvable Lie algebra, a subcase being nilpotent algebras. One can define the one-forms $\lbrace e^a \rbrace$, dual to the algebra vectors; they satisfy the relation
\beq
\d e^a= - \frac{1}{2} f^a{}_{bc} e^b \w e^c \ . \label{de}
\eeq
For all cases considered in this paper, the Lie algebra and Lie group are in one-to-one correspondence; in particular the solvable algebras will be almost abelian. The one-forms and relation \eqref{de} can then be promoted to the cotangent bundle of the corresponding Lie group, the latter viewed as a manifold. To get the solvmanifold, the group should further be divided by a lattice: this action makes discrete identifications that result in particular in having a compact manifold. The existence of a lattice is not always guaranteed (see e.g. \cite{Andriot:2010ju}), although it exists for all manifolds considered in this paper. The choice of the lattice can also impact the solvmanifold cohomology, as we will come back to.

The one-forms $e^a$ are globally defined on the final solvmanifold, i.e. they are invariant under the lattice discrete identifications. They are typically understood as the Maurer-Cartan one-forms, given locally in terms of the vielbein $e^a{}_m$ by $e^a=e^a{}_m \d y^m$ (with the flat metric denoted $\eta_{ab}$). The dual vectors are then given by $\del_a = e^m{}_a \del_m = E_a$, and the structure constants by $f^a{}_{bc}= 2 e^a{}_m \del_{[b} e^m{}_{c]}$. A necessary requirement for compactness is unimodularity of the algebra: $\sum_a f^a{}_{ab}=0$; this will be satisfied in the cases studied here. The spin connection for Levi-Civita connection can be expressed in terms of the structure constants; the same holds for the Ricci tensor ${\cal R}_{ab}$ and scalar ${\cal R}$ (see e.g. \cite{Andriot:2014uda}). For constant $f^a{}_{bc}$ as here, and provided unimodularity, one has (in flat indices)
\bea
& 2\ {\cal R}_{cd} =  - f^b{}_{ac} f^a{}_{bd} - \eta^{bg} \eta_{ah} f^h{}_{gc} f^a{}_{bd} + \frac{1}{2} \eta^{ah}\eta^{bj}\eta_{ci}\eta_{dg} f^i{}_{aj} f^g{}_{hb} \ ,\label{Ricciflat}\\
& 2\ {\cal R}=  - \eta^{cd} f^a{}_{bc} f^b{}_{ad} -\frac{1}{2} \eta_{ad} \eta^{be} \eta^{cg} f^a{}_{bc} f^d{}_{eg}  \ .\label{Ricflat}
\eea

For physics purposes and future convenience, we introduce slightly different notations. From now on, the one-forms $e^a$ are given in terms of a real vielbein matrix $e$ such that $|e|\equiv |{\rm det}(e)|=1$. This means that the radii, warp factor, and further diagonal contributions to the metric are not contained in $e$ but in a diagonal metric denoted $g_{ab}$
\beq
g_{mn}=e^a{}_m ~ g_{ab} ~ e^b{}_n \ . \label{defmetric}
\eeq
The $f^a{}_{bc}$ to be used are still those of \eqref{de}: they encode purely the fibration structure of the manifold, without any conformal scaling. To get the structure constants associated to a constant $g_{ab}$, each index of $f^a{}_{bc}$ should be rescaled by a diagonal $\sqrt{g_{aa}}$; similar rescalings should be made in \eqref{Ricciflat} and \eqref{Ricflat}, while the unimodularity condition is not modified in this new notation. If $g_{ab}$ is not constant, further changes occur, but we do not need to work them out as we will mostly use the smeared metric $\tilde{g}_{ab}$, which will be constant
\beq
\tilde{g}_{ab} \equiv g_{ab}|_{A=0} \ .\label{tildemetric}
\eeq

We finally introduce the (six-dimensional) Laplacian of a function $\varphi$
\beq
\Delta \varphi = g^{mn} \nabla_m \del_n \varphi = \frac{1}{\sqrt{|g|}} \del_m (\sqrt{|g|} g^{mn} \del_n \varphi ) = \frac{1}{\sqrt{|g|}} \del_m ( e^m{}_a \sqrt{|g|} g^{ab} \del_b \varphi ) \ .
\eeq
The unimodularity condition is expressed in terms of the vielbeins as
\beq
\!\!\!\!\! 0=f^a{}_{ab}=  \del_{m} e^m{}_{b} + e^n{}_{b} e^m{}_{a} \del_{n} e^a{}_m = \del_{m} e^m{}_{b} + e^n{}_{b} {\rm Tr} (e^{-1} \del_{n} e) = \del_{m} e^m{}_{b} + e^n{}_{b} \del_{n} \ln ({\rm det} ~ e) \ .
\eeq
In our notations, ${\rm det} ~ e$ is constant, implying $\del_{m} e^m{}_{b} =0$ and then
\beq
\Delta \varphi = \frac{1}{\sqrt{|g|}} \del_a ( \sqrt{|g|} g^{ab} \del_b \varphi ) \ . \label{Lapl}
\eeq
The $e^a{}_m$ do not contain any warp factor in our notations, so we can follow the same reasoning for the smeared metric $\tilde{g}_{ab}$ to obtain the associated Laplacian
\beq
\tilde{\Delta} \varphi = \frac{1}{\sqrt{|\tilde{g}|}} \del_a ( \sqrt{|\tilde{g}|} \tilde{g}^{ab} \del_b \varphi ) \ . \label{tLapl}
\eeq
We recall $\del_a = e^m{}_a \del_m = E_a$: these will appear through the exterior derivative $\d= \d y^m \del_m= e^a \del_a$. One should keep in mind that they are not bare coordinate derivatives, but typically contain combinations of derivatives due to the inverse vielbein. It will be clear in the examples.

\subsection{Ricci flat solvmanifolds}\label{sec:flattwtor}

Because of physics motivations, we are interested in solvmanifolds that admit a Ricci flat metric. Note that such a property holds equivalently at the level of the solvable group, since this is a local statement, and the lattice action does not bring any singularity. Among all six-dimensional solvable algebras (a typical number is obtained by counting indecomposable ones: there are 164 of them), only three lead to Ricci flat solvmanifolds, on top of the trivial algebra giving the torus. This surprising result was proven in \cite{Grana:2013ila} even beyond solvable groups, for Lie group manifolds (see also references therein). Before discussing these three algebras \eqref{s1s2s3} and manifolds, we give in the following paragraph a simple argument to reproduce and explain this result.

A Ricci flat solvmanifold should have at least a vanishing Ricci scalar \eqref{Ricflat}
\beq
2\ {\cal R}=  - \tilde{g}^{cd} f^a{}_{bc} f^b{}_{ad} -\frac{1}{2} \tilde{g}_{ad} \tilde{g}^{be} \tilde{g}^{cg} f^a{}_{bc} f^d{}_{eg}  = 0 \ , \label{Rdiscuss}
\eeq
using notations introduced in section \ref{sec:twtor}: $\tilde{g}_{ab}$ is diagonal and contains the radii, it is considered constant here, while the structure constants only capture the topological information of the fibration. Solvmanifolds were argued in \cite{Andriot:2010ju} to allow only for three types of fibrations: those of nilmanifolds, the ``hyperbolic'' and the ``standard'' rotations, each of them corresponding to specific structure constants. In nilmanifolds, there is a definite hierarchy in the fibration, equivalently to the algebra: it results in the fact that one cannot have both $f^a{}_{bc}$ and $f^b{}_{ad}$ non-zero, even for $a=b$. Consequently, the first term of \eqref{Rdiscuss} vanishes for a nilmanifold. The second term is in any case a sum of squares, so it is strictly negative (see also \cite{Ngaha} and references therein). On the contrary, fibrations given by hyperbolic and standard rotations lead to $f^a{}_{bc}$ and $f^b{}_{ad}$ together non-zero: more precisely, given a fixed $c$ and some pair of directions $a\neq b$, one gets $f^a{}_{bc} f^b{}_{ac} >0$ for the hyperbolic case and $f^a{}_{bc} f^b{}_{ac}<0$ for the standard one.\footnote{These structure constants correspond to vielbeins made of $\cosh, \sinh$ for ``hyperbolic'' rotations, and $\cos, \sin$ for ``standard'' ones (see e.g. \eqref{vielbeins3}), hence the names chosen. The former case can be reparameterized in terms of exponentials, corresponding to structure constants of the form $f^a{}_{ac}\neq 0$ for a given $a$; this parametrization however does not provide globally well-defined one-forms \cite{Andriot:2010ju}.} As a consequence, the first term of \eqref{Rdiscuss} is negative for the former fibrations, but positive for the latter. Then, only for standard rotations, the first term has a chance to cancel the second one: by appropriately counting the indices, it can be verified to occur for specific values of structure constants with respect to the radii. In more details, fixing this relative freedom by taking, for one pair of directions $a\neq b$, the structure constants equal in absolute value, ${\cal R}$ vanishes for the pair of radii being equal (see the explicit computation \eqref{Riccitensors3}). Taking the pair of radii to be different brings the Ricci scalar back to a negative value, meaning that this fibration never leads to a strictly positive ${\cal R}$. There is thus no possible ``excess'' that could cancel negative contributions from other fibrations, i.e. any combination with another type of fibration, as in a random solvmanifold, would lead to a negative Ricci scalar. We conclude that only solvmanifolds with fibrations made purely of standard rotations can satisfy \eqref{Rdiscuss} (the pairs of radii should then be equal), and thus have a chance to be Ricci flat. As detailed below, only three algebras have structure constants corresponding to such fibrations: we recover this way the result of \cite{Grana:2013ila}.

Ricci flat solvmanifolds have a specific fibration structure: it consists in pairs of directions $a\neq b$ fibered with standard rotations over $c$, corresponding to pairs of non-zero structure constants $f^a{}_{bc}, f^b{}_{ac}$ whose product is negative. In six dimensions, up to relabeling, only three such solvable algebras can be constructed: with one or two pairs, and for the latter, fibered over different or the same directions. This leads respectively to
\bea
s_1&: \d e^1= q~ e^2 \w e^3 \ ,\ \d e^2= -q~ e^1 \w e^3 \ ,\ \d e^3=\d e^4=\d e^5=\d e^6= 0 \ ,\label{s1s2s3}\\
s_2&: \d e^1= q~ e^2 \w e^3 \ ,\ \d e^2= -q~ e^1 \w e^3 \ ,\ \d e^4= rq~ e^5 \w e^6 \ ,\ \d e^5= - rq~ e^4 \w e^6\ ,\ \d e^3=\d e^6= 0 \ ,\nn\\
s_3&: \d e^1= q~ e^2 \w e^5 \ ,\ \d e^2= -q~ e^1 \w e^5 \ ,\ \d e^3= rq~ e^4 \w e^5 \ ,\ \d e^4= - rq~ e^3 \w e^5\ ,\ \d e^5=\d e^6= 0 \ ,\nn
\eea
with for now $q$ and $r \in \mathbb{R}^*$.\footnote{The parameter $q$ can be reabsorbed by rescaling so it does not provide inequivalent algebras; it will however be crucial when choosing the lattice as discussed below.} The algebras corresponding to $s_1$ and $s_2$ are denoted ISO(2)$\times$U(1)$^3$ and ISO(2)$\times$ISO(2) in \cite{Grana:2013ila}; $s_1$, $s_2$, $s_3$ are denoted s4.1, s2.4, s2.5 in \cite{Grana:2006kf}; in the math literature, algebras corresponding to $s_1$, $s_2$, and $s_3$ with $r^2=1$, are denoted $\mathfrak{e}(2) \oplus \mathbb{R}^3$, $\mathfrak{e}(2) \oplus \mathfrak{e}(2)$, $\mathfrak{g}_{5.17}^{0,0,\pm1} \oplus \mathbb{R}$, with $\mathfrak{e}(2)=\mathfrak{g}_{3.5}^0$ \cite{Bock}. Note that when $r^2=1$ in $s_3$, we know that a lattice exists (for the algebra in a more general case, see e.g. footnote 8 in \cite{Andriot:2010ju}); in addition, we will obtain vacua only in that case.

The algebra corresponding to $s_3$ is given as in \eqref{algebra} by the non-zero commutators
\beq
\lbrack E_2 , E_5 \rbrack = -q~ E_1\ ,\ \lbrack E_1 , E_5 \rbrack = q~ E_2\ ,\ \lbrack E_4 , E_5 \rbrack = - rq~ E_3 \ ,\ \lbrack E_3 , E_5 \rbrack = rq~ E_4 \ . \label{algs3}
\eeq
One has $E_a = \del_a = e^m{}_a \del_m= (e^{-T} \del)_a$. The one-forms are given by \cite{Andriot:2010ju}
\bea
& e^5=\d y^5,\ e^6=\d y^6, \label{vielbeins3} \\
& \begin{pmatrix} e^1 \\ e^2 \end{pmatrix} = \begin{pmatrix} \cos (q y^5) & -\sin (q y^5) \\ \sin (q y^5) & \cos (q y^5) \end{pmatrix} \begin{pmatrix} \d y^1 \\ \d y^2 \end{pmatrix},\ \begin{pmatrix} e^3 \\ e^4 \end{pmatrix} = \begin{pmatrix} \cos (rq y^5) & -\sin (rq y^5) \\ \sin (rq y^5) & \cos (rq y^5) \end{pmatrix} \begin{pmatrix} \d y^3 \\ \d y^4 \end{pmatrix} \ ,\nn
\eea
where one reads-off the vielbein matrix, made of the standard rotations discussed above. One can verify ${\rm det} ~ e=1$ in agreement with notations of section \ref{sec:twtor}; this holds as well for $s_1$ and $s_2$. Finally, we compute the Ricci tensor \eqref{Ricciflat} using as in \eqref{Rdiscuss} the smeared metric
\bea
2 {\cal R}_{ab} &= \delta^1_a \delta^1_b ~ q^2 \tilde{g}^{55} \tilde{g}^{22} \left( (\tilde{g}_{11})^2 - (\tilde{g}_{22})^2 \right) + \delta^2_a \delta^2_b ~ q^2 \tilde{g}^{55} \tilde{g}^{11} \left( (\tilde{g}_{22})^2 - (\tilde{g}_{11})^2 \right) \label{Riccitensors3}\\
& + \delta^3_a \delta^3_b ~ r^2 q^2 \tilde{g}^{55} \tilde{g}^{44} \left( (\tilde{g}_{33})^2 - (\tilde{g}_{44})^2 \right) + \delta^4_a \delta^4_b ~ r^2 q^2 \tilde{g}^{55} \tilde{g}^{33} \left( (\tilde{g}_{44})^2 - (\tilde{g}_{33})^2 \right) \nn \\
& + \delta^5_a \delta^5_b ~ \left( q^2 \left( 2- \frac{\tilde{g}_{11}}{\tilde{g}_{22}} - \frac{\tilde{g}_{22}}{\tilde{g}_{11}} \right) + r^2 q^2 \left( 2- \frac{\tilde{g}_{33}}{\tilde{g}_{44}} - \frac{\tilde{g}_{44}}{\tilde{g}_{33}} \right) \right) \ . \nn
\eea
The metric is Ricci flat if and only if the pairs of radii are equal, i.e. $\tilde{g}_{11} = \tilde{g}_{22}$ and $\tilde{g}_{33}=\tilde{g}_{44}$; a condition already obtained above from ${\cal R}$. The manifold is then indeed Ricci flat!\\

Although unnecessary so far, let us discuss the choice of the lattice, that takes the solvable group to the solvmanifold. The algebras corresponding to $s_1$, $s_2$, $s_3$ are ``not completely solvable'': they can lead to inhomogeneous manifolds, and different choices of lattice can give different solvmanifolds with their own topology and cohomology, thus different physics. Some lattices even lead to a torus as we will see. For a ``completely solvable'' algebra (also nilpotent ones), there exists an isomorphism between the algebra cohomology and that of the solvmanifold \cite{Bock}, but this is not guaranteed otherwise, as here. In our conventions, coordinates $y^m$ are angles: for $s_3$, the points $y^5$ and $y^5 + 2\pi$ should then be identified by the lattice to get a circle, while identifications should be made accordingly among the pair $y^1, y^2$, and the pair $y^3, y^4$, so that the $e^a$ remain globally defined. Those lattice identifications are achieved by two dimensional rotations involving $\cos(q 2\pi)$ or $\cos(rq 2\pi)$, etc., making $q$ and $r$ crucial parameters in the lattice choice. If $q$ and $r$ are integers, identifications of the pairs of coordinates are trivial: the $2\pi$ rotations are identities and the coordinates are identified with themselves. All the forms $\d y^m$ are then globally defined, and this results in having a torus \cite{Grana:2013ila}. Choosing another value for $q$ can lead to less trivial coordinate identifications and non-globally defined $\d y^m$, while still having globally defined $e^a$. The different cases and cohomologies were actually worked-out for $s_1$ and $s_3$ in \cite{Console:2012wp} (see in particular table 1).\footnote{We believe there is a typo in table 1 of \cite{Console:2012wp}. For $G^{p,-p,r}_{5.17}\times \mathbb{R}$ with $\ov{t}=2\pi r_2$ (where $r_2$ is meant as a non-zero integer), the two lines in the solvmanifold cohomology seem to be exchanged: one should have the torus cohomology ($b_1=6$, etc.) for $p=0$, and the other one ($b_1=2$, etc.) for $p\neq0$, as discussed there in the corresponding text.} For $s_3$ with $r=\pm 1$, the algebra cohomology is given by $b_1=2, b_2=5, b_3=8$, but different choices of the lattice allow for three different possible cohomologies of the solvmanifold
\bea
q \in \mathbb{Z}^*:&\ b_1=6, b_2=15, b_3=20 \ ,\label{cohomo}\\
q=\frac{1}{2}:&\ b_1=2, b_2=7, b_3=12 \ ,\nn\\
q=\frac{1}{4}:&\ b_1=2, b_2=5, b_3=8 \ ,\nn
\eea
the first one corresponding to the torus as discussed above, while the last one matches the algebra cohomology. For $s_1$, there are two possible cohomologies: that of the torus, or that of the algebra (obtained by different lattices) given by $b_1=4, b_2=7, b_3=8$. Finally, note that the Euler characteristic vanishes, $\chi (\mmm) =0$, in all cases. We refer the reader to \cite{Bock, Console:2012wp, Grana:2013ila} and references therein for more details on this topic; in particular a classification of lattices can be found in \cite{Grana:2013ila}. Further techniques to compute cohomologies of solvmanifolds are presented in \cite{Console, Daniele}, although they are not used on examples of interest here. Understanding the role of the lattice will be important in sections \ref{sec:Td} and \ref{sec:CY}. To present our vacua though, we will not need in this paper to specify explicitly the value of $q$, so we simply keep in mind that an appropriate lattice choice can give from $s_1$, $s_2$, $s_3$ solvmanifolds that are not a torus.

\subsection{An account of supersymmetric Minkowski flux vacua on solvmanifolds}\label{sec:soltwistedtori}

The conditions to find supersymmetric Minkowski flux vacua have been presented in section \ref{sec:susysol} (see the summary at the end). Let us discuss here the solutions to these conditions for internal manifolds being solvmanifolds (introduced in section \ref{sec:twtor}). The SUSY condition on $\Phi_1$ \eqref{SUSY} gives the requirement of admitting a closed pure spinor, up to the $H$-flux that only brings a twist: in GCG terms, this characterises the underlying manifold as being a Generalized Calabi-Yau (GCY) \cite{Grana:2004bg}. All six-dimensional nilmanifolds have actually been shown to be GCY \cite{CG}: this makes them, and more generally solvmanifolds, interesting candidates to look for solutions, on top of being in practice easy to handle. A scan for solutions on all six-dimensional nilmanifolds and some solvmanifolds was performed in the seminal paper \cite{Grana:2006kf}. Despite this huge analysis lead to quite some solutions, it remains incomplete as we now explain.

\begin{itemize}
\item Only solutions with SU(3) or orthogonal SU(2) structure were looked for. The possibility of an intermediate SU(2) structure was realised later; such solutions were obtained in \cite{Koerber:2007hd, Andriot:2008va} (see also \cite{Andriot:2009rc}). Other types of solutions are a priori possible when letting coefficients vary (see section 4.3 of \cite{Andriot:2014qla} for more details and references on those): in particular, so-called dynamical SU(2) structure would have spinors varying on the manifold, possibly merging at some points (these could correspond to ``type changing loci''); no such solution has been found so far (with a compact $\mmm$).

\item Only solvmanifolds where $f^a{}_{ab}=0$ {\it without sum} were considered. This leaves aside many others, whose fibration involves ``hyperbolic'' rotations (see section \ref{sec:flattwtor}). Solutions were found on these other solvmanifolds: one in \cite{Camara:2005dc} and one in \cite{Andriot:2010ju} (see \cite{Andriot:2010sya} for a list).

\item The method used in \cite{Grana:2006kf} leaves room to more localized solutions. It consists in first finding solutions in the smeared limit, and then trying to localize them, including in particular warp factors. Two types of smeared solutions at least are then problematic: those with intersecting sources, and the fluxless ones. Localizing the former is a famous issue in supergravity (see \cite{Smith:2002wn} for a review and e.g. \cite{Arapoglu:2003ah} for more references), not overcome in \cite{Grana:2006kf}; for the latter, a purely localized flux cannot be recovered with a simple multiplication by warp factors, on the contrary to smeared solutions with fluxes. We come back in details to this point in section \ref{sec:propsources}, while obtaining new localized solutions, fluxless in the smeared limit, in section \ref{sec:newsol}. The only previously known example of such a solution on a solvmanifold was given in section 2.3.1 of \cite{Andriot:2010ju}.

\item The analysis of \cite{Grana:2006kf} was aiming at an exhaustive search for vacua with given criteria. Within that set, some solutions may still have been missed: we present one such vacuum in section \ref{sec:sol2}; a smeared version of the latter was given in \cite{Kachru:2002sk}.

\end{itemize}

\noindent In addition to the list of solutions given so far, let us mention the possibility of bare Ricci flat solvmanifolds: the metric with constant warp factor and dilaton, without any flux and source, satisfies the equations of motion. The solvmanifolds for such simple vacua were presented in section \ref{sec:flattwtor}, and the corresponding three algebras \eqref{s1s2s3} were already identified in \cite{Grana:2006kf} when discussing fluxless solutions. As addressed there, whether these vacua are SUSY should be verified and depends only on geometric properties of the manifold; we come back to this point in sections \ref{sec:newsol} and \ref{sec:CY}. Such a simple vacuum on a solvmanifold corresponding to $s_1$ was used in \cite{Hassler:2014sba} to get T-dual solutions.  Thanks to their specificities, these vacua do not require $O$-planes and $D$-branes, but these sources can still be considered present, either on top of each other, or at different places but smeared; the solutions of section \ref{sec:newsol} will localize the latter. The former was the setting for the fluxless vacua of \cite{Grana:2006kf}, where the orientifold projection was then taken into account.

We now mention further related references. A geometrical characterisation of the SUSY conditions for an intermediate SU(2) structure was obtained in \cite{Fino:2010dx}, and some solvmanifolds satisfying those were given. Solutions on solvmanifolds also appeared as examples when studying generalizations of mirror symmetry \cite{Grana:2006kf, Lau}. Finally, other Lie algebra based manifolds have been considered: a general presentation is given in \cite{Danielsson:2011au}; see \cite{Lust:1986ix} for semi-simple algebras and six-dimensional cosets, and \cite{Koerber:2008rx} for further suitable cosets.\\

Given this list of solutions, one should study their ``novelty'' as raised in \cite{Grana:2006kf}, meaning whether they are T-dual to (geometric) solutions on a torus, as first obtained in \cite{Kachru:2002sk}. T-dual vacua would lead to the same effective theory after compactification, up to four-dimensional field redefinitions, so a T-dual solution would not lead to new physics. Subtleties on the lattice choice and cohomologies discussed in section \ref{sec:flattwtor} should also be kept in mind when compactifying and studying this question of the novelty \cite{Grana:2013ila}: one should not end-up on the torus. The only ``truly new'' solutions (not T-dual to one on a torus) in \cite{Grana:2006kf} were involving smeared intersecting sources, and could not be localized. The same goes for solutions in \cite{Camara:2005dc, Andriot:2008va, Andriot:2010ju}, except for the one in section 2.3.1 of \cite{Andriot:2010ju}. The latter could then be the only localized SUSY Minkowski flux vacuum of type II supergravities on a solvmanifold, that is not T-dual to one on a torus. We come back to it in details and give further examples in section \ref{sec:newsol}, and prove in section \ref{sec:Td} they are all indeed truly new vacua.

\section{Warming-up: vacua on nilmanifolds, and beyond}\label{sec:solnil}

We present in this section two SUSY Minkowski flux vacua of type IIA supergravity where the internal manifold is a nilmanifold. A smeared version of both vacua was given in \cite{Kachru:2002sk}. The first one allows us to fix the notations; note that the second one does not appear in \cite{Grana:2006kf}. Building on these two examples, we then discuss general properties related to subspaces wrapped by the sources, allowing us to go beyond the vacua found so far, and obtain new ones on solvmanifolds as presented in section \ref{sec:newsol}.\\

The nilmanifolds of the two vacua both involve the only (non-trivial) three-dimensional nilmanifold $N_3$. The four-dimensional extension of the latter by a simple product with a circle is also known as the Kodaira-Thurston manifold. $N_3$ is built out of the Heisenberg algebra, expressed in terms of one-forms \eqref{de} as
\beq
\d e^1=0 \ ,\ \d e^2=0\ ,\ \d e^3 = f e^1\w e^2 \ ,\label{e123}
\eeq
with structure constant $f^3{}_{12}=-f \neq 0$. One expression of the one-forms is
\beq
e^1=\d y^1\ , \ e^2=\d y^2\ , \ e^3=\d y^3 + f y^1 \d y^2 \ , \label{Heisea}
\eeq
where one verifies $|e|=1$ as required in section \ref{sec:twtor}. The lattice action, i.e. the discrete identifications of coordinates that leave the $e^a$ invariant, can be read from \eqref{Heisea}.

Being the simplest nilmanifold besides the three-dimensional torus $T^3$, the ``Heisenberg manifold'' $N_3$ appeared many times in the string theory literature. SUSY Minkowski flux vacua on $\mmm=N_3 \times T^3$ (algebra denoted n5.2) and on $N_3 \times N_3$ (algebra denoted n4.5) are mentioned in \cite{Grana:2006kf}. The former is with an $O_4$ and orthogonal SU(2) structure, while the latter allows for two vacua with an $O_5$, one with SU(3) and one with orthogonal SU(2) structure.\footnote{A vacuum on $N_3 \times N_3$ with an $O_5$ and intermediate SU(2) structure is then likely to exist, as e.g. in \cite{Andriot:2008va}.} These vacua being T-dual to others on a torus $T^6$, they are not given explicitly in \cite{Grana:2006kf}. Information on the one with SU(3) structure on $N_3 \times N_3$ can be found in \cite{Andriot:2010ju}, while the vacuum on $N_3 \times T^3$ is detailed below in section \ref{sec:sol1}. This vacuum was first mentioned in \cite{Kachru:2002sk}, after a T-duality from $T^6$. A further T-duality was shown to give a non-geometric background, as studied in more details in \cite{Marchesano:2007vw, Andriot:2011uh}. This vacuum is thus important in the non-geometry literature, where it is sometimes referred to as the ``toroidal example''. Closed string was partially quantized on it in \cite{Andriot:2012vb}. Finally, $N_3$ appeared in a stringy inflation scenario making use of its monodromy properties \cite{Silverstein:2008sg}; this will be the topic of \cite{Andriot:2015aza}.\\

The conditions to get SUSY Minkowski flux vacua were summarized into a list at the end of section \ref{sec:susysol}, and made more explicit in appendix \ref{ap:condSUSYvac}; the vacua presented below are solutions to all these constraints. In addition, we will use the following convenient parametrization of the structure forms. For an orthogonal SU(2) structure, we will consider
\beq
z=\sqrt{t_3}~ z^3 \ ,\ \omega=\sqrt{t_1t_2}~ z^1 \w z^2\ ,\ j=\frac{\i}{2} (t_1~ z^1\w \ov{z^1} + t_2~ z^2\w \ov{z^2})\ ,\ t_i>0 \ , \label{SU2forms}
\eeq
different than in \cite{Grana:2006kf, Andriot:2010sya} but equivalent. The $t_i$ are constant. The complex one-forms $z^i$ will take generically the form $z^1=e^{\pm A} e^1 + \i e^{\pm A} \tau e^2$, etc., where the $\tau$ are non-zero real constants, and one has $e^{+A} e^a$ when the source is along $e^a$, or $e^{-A} e^a$ when it is transverse to it. The $z^i$ will provide a basis: there will be a one-to-one map to the $e^a$. The parametrization \eqref{SU2forms} then makes the compatibility conditions \eqref{compatjo} and \eqref{compatz} automatically satisfied, especially the contractions thanks to the use of a diagonal metric. In addition, the $z^i$ will be taken as (1,0)-forms with respect to the almost complex structure $I$. With $J\equiv j + \frac{\i}{2} z \w \ov{z}$, the Hermitian metric is then obtained as $g_{i\ov{j}}=-I_i{}^k J_{k\ov{j}}$, $g_{\ov{i}j}=-I_{\ov{i}}{}^{\ov{k}} J_{\ov{k}j}$. Changing basis from $z^i$ to $e^a$, we deduce $g_{ab}$ \eqref{defmetric}, given generically by
\beq
g_{ab}={\rm diag} (t_1 e^{\pm 2A},\ t_1 e^{\pm 2A} \tau^2,\ \dots ) \ .
\eeq
It is definite-positive as it should be, and the warp factor has the right powers. We will use a similar parametrization of the SU(3) structure forms
\beq
J = \frac{\i}{2} \left( t_1~ z^1 \w \ov{z}^1 + t_2~ z^2 \w \ov{z}^2 + t_3~ z^3 \w \ov{z}^3 \right) \ ,\ \Omega= \sqrt{t_1 t_2 t_3}~ z^1 \w z^2 \w z^3 \ ,\ t_{1,2,3} > 0 \ , \label{SU3forms}
\eeq
making the compatibility conditions \eqref{compatSU3} automatically satisfied. Once again, the basis of (1,0)-forms $z^i$ will allow to build the corresponding metric $g_{ab}$. We are now ready to present the vacua.

\subsection{Vacuum with $O_4$}\label{sec:sol1}

We present here explicitly the vacuum on $N_3\times T^3$ mentioned above. The torus directions complete \eqref{e123} with
\beq
\d e^4=\d e^5=\d e^6 = 0 \ .\label{e456}
\eeq
The solution to the constraints has an orthogonal SU(2) structure, and an $O_4$ along $e^3$. The dilaton is taken to be $e^{\phi}=g_s e^A $ with a constant $g_s$. The structure forms are given by \eqref{SU2forms} with
\beq
z^3= e^{-A} e^6 + \i e^A \tau_3 e^3 \ ,\ z^1=e^{-A} e^1 + \i e^{-A} \tau_2 e^2\ ,\ z^2=e^{-A} e^4 + \i e^{-A} \tau_5 e^5 \ .
\eeq
We deduce the metric
\beq
g_{ab}=\mbox{diag}(e^{-2A} t_1, e^{-2A} t_1 (\tau_2)^2, e^{2A} t_3 (\tau_3)^2, e^{-2A} t_2, e^{-2A} t_2 (\tau_5)^2, e^{-2A} t_3 ) \ .\label{metricsolN3}
\eeq
In addition, $H=0$ and the only non-zero RR flux is (from \eqref{SUSYSU2})
\bea
F_4 =g_s^{-1} \sqrt{t_3}\tau_3 & \left( e^{-4A} *\left( \d(e^{4A})\w e^3\right) + f \frac{\sqrt{|g|}}{g_{11}g_{22}} e^3\w e^4\w e^5\w e^6 \right) \\
=g_s^{-1} \frac{\tau_3 \sqrt{|g|}}{|\tau_3| e^{5A} \sqrt{g_{33}}} &\Bigg( - g^{11} \del_1 (e^{4A}) e^2\w e^4\w e^5\w e^6  + g^{22} \del_2 (e^{4A}) e^1\w e^4\w e^5\w e^6  \nn\\
&\ - g^{44} \del_4 (e^{4A}) e^1\w e^2\w e^5\w e^6 + g^{55} \del_5 (e^{4A}) e^1\w e^2\w e^4\w e^6 \nn\\
&\ - g^{66} \del_6 (e^{4A}) e^1\w e^2\w e^4\w e^5 + e^{4A} f \frac{g_{33}}{g_{11}g_{22}} e^3\w e^4\w e^5\w e^6 \Bigg) \label{F40sol1}
\eea
using the Hodge star \eqref{Hodge} and the diagonal metric $g_{ab}$. We recall from section \ref{sec:twtor} that the $\del_a=E_a$ appearing above differ from the bare coordinate derivatives by a factor $e^m{}_a$. Here in practice, the only difference is for $\del_{a=2}=\del_{y^2} - f y^1 \del_{y^3}$ deduced from \eqref{Heisea}. The $O_4$ being along $e^3$, $A$ is taken independent of $y^3$. One can thus consider $\del_m$ instead of $\del_a$ in the $F_4$ expression. Such a simplification will not happen for the vacua of section \ref{sec:newsol}. Using the metric without warp factor (or smeared) \eqref{tildemetric}, we rewrite $F_4$ as
\bea
F_4 = - g_s^{-1} \frac{\tau_3 \sqrt{|\tilde{g}|}}{|\tau_3| \sqrt{\tilde{g}_{33}}} &\Bigg( - \tilde{g}^{11} \del_1 (e^{-4A}) e^2\w e^4\w e^5\w e^6  + \tilde{g}^{22} \del_2 (e^{-4A}) e^1\w e^4\w e^5\w e^6  \nn\\
&\ - \tilde{g}^{44} \del_4 (e^{-4A}) e^1\w e^2\w e^5\w e^6 + \tilde{g}^{55} \del_5 (e^{-4A}) e^1\w e^2\w e^4\w e^6 \nn\\
&\ - \tilde{g}^{66} \del_6 (e^{-4A}) e^1\w e^2\w e^4\w e^5 - f \frac{\tilde{g}_{33}}{\tilde{g}_{11}\tilde{g}_{22}} e^3\w e^4\w e^5\w e^6 \Bigg) \ . \label{Fsol1}
\eea
This makes it simpler to get $\d F_4$, expressed in terms of the unwarped Laplacian \eqref{tLapl}
\beq
\d F_4=g_s^{-1} \frac{\tau_3}{|\tau_3|} \left( \tilde{\Delta} (e^{-4A}) +  f^2 \frac{\tilde{g}_{33}}{\tilde{g}_{11}\tilde{g}_{22}} \right) \sqrt{\frac{|\tilde{g}|}{\tilde{g}_{33}}}\ e^1\w e^2\w e^4\w e^5\w e^6 \ .\label{BI1sol1}
\eeq
As the other constraints, the calibration condition \eqref{calib} is satisfied (phases are fixed as in appendix \ref{ap:condSUSYvac}): we get ${\rm vol}_{||}=\sqrt{g_{33}} e^3$ upon fixing the sign $\tau_3 >0$; this sign could though be relaxed and understood as an orientation. The convention \eqref{defvolbot} then gives the volume transverse to the source
\beq
{\rm vol}_{\bot}= - \sqrt{\frac{|g|}{g_{33}}} \ e^1\w e^2\w e^4\w e^5\w e^6 \ ,
\eeq
and we rewrite \eqref{BI1sol1} as
\beq
\d F_4= - g_s^{-1}  \left( \tilde{\Delta} (e^{-4A}) +  f^2 \frac{\tilde{g}_{33}}{\tilde{g}_{11}\tilde{g}_{22}} \right) e^{5A} {\rm vol}_{\bot} \ .\label{BI2sol1}
\eeq
As discussed at the end of appendix \ref{ap:condSUSYvac}, \eqref{BI2sol1} is precisely compatible with the general BI \eqref{BI}, where the sources, along $e^3$, are the $O_4$ and possibly some $D_4$. As usual, the unwarped Laplacian would generate the $\delta$ localizing the sources, up to a standard shift in $f^2$, a contribution from the geometric flux (or curvature) of $N_3$. This term is the T-dual counterpart to $H\w F_3$ (coming from an ISD flux) shifting $\d F_5$ in a torus vacuum with $O_3$, as in \cite{Giddings:2001yu}. In both cases, the shift term contributes to the tadpole cancelation, and resulting quantization condition that we refrain from working-out here. Note that the metric components multiplying $f^2$ are precisely those needed to rescale each index of $f^3{}_{12}=-f$ into the full structure constant, as explained below \eqref{defmetric}. Finally, the shift term can also be understood as due to the fact that sources do not wrap a cycle, but a generalized cycle; we come back to this in section \ref{sec:propsources}.

\subsection{Vacuum with $O_6$}\label{sec:sol2}

We now present a vacuum on a nilmanifold (particular example of a solvmanifold) made of a five-dimensional manifold times a free circle. The fibration in the former is twice that of the Heisenberg manifold, although fibered over the same circle along $e^2$. The corresponding algebra, denoted n4.6 in \cite{Grana:2006kf}, is expressed in terms of one-forms as
\beq
\d e^1 = \d e^2 = \d e^4 = \d e^5 = 0 \ ,\ \d e^3 = f_1 ~ e^1 \w e^2 \ ,\ \d e^6 = f_2 ~ e^2 \w e^5 \ ,
\eeq
with non-zero structure constants $f^3{}_{12}=-f_1$, $f^6{}_{25}=-f_2$. The vacuum has an $O_6$ along $e^3 \w e^4 \w e^6$.  A smeared version of this vacuum was given in \cite{Kachru:2002sk}. It is however not mentioned in \cite{Grana:2006kf}, so we believe it has been missed there. The only vacua in \cite{Grana:2006kf} on this manifold have an $O_5$ along $e^3 \w e^6$: those are very likely T-dual to ours along the free circle $e^4$. Both are also very probably T-dual to vacua on $T^6$ \cite{Kachru:2002sk}, which makes our vacuum not ``truly new'', as discussed in section \ref{sec:soltwistedtori}. On top of the $O_6$, our vacuum has an orthogonal SU(2) structure: it is the first localized SUSY Minkowski flux vacuum with both attributes on a solvmanifold; we will present another one in section \ref{sec:newsol}.

The solution to all constraints is first given by the SU(2) structure forms \eqref{SU2forms} with
\beq
z^3= e^{A} e^4 + \i e^{-A} \tau_3 e^2 \ ,\ z^1=e^{-A} e^1 + \i e^{-A} \tau_1 e^5\ ,\ z^2=e^{A} e^3 + \i e^{A} \tau_2 e^6 \ .
\eeq
The corresponding metric is
\beq
g_{ab}=\mbox{diag}(e^{-2A} t_1, e^{-2A} t_3 (\tau_3)^2, e^{2A} t_2, e^{2A} t_3, e^{-2A} t_1 (\tau_1)^2, e^{2A} t_2 (\tau_2)^2 ) \ .\label{metricsolN5}
\eeq
The dilaton is $e^{\phi}= g_s e^{3A}$ with a constant $g_s$. We take $A(y^1,y^2,y^5)$. The SUSY conditions \eqref{SUSYSU2} are satisfied with one constraint
\beq
\frac{\tau_1}{\tau_2} = - \frac{f_2}{f_1} \ , \label{constraintsol2}
\eeq
that we now consider verified. In addition, $H=0$ (due to its BI) and the only non-zero RR flux is
\bea
F_2= - g_s^{-1} \sqrt{t_3} t_2 \tau_2 & \Bigg( e^{-4A} *(\d (e^{4A})\w e^3 \w e^4 \w e^6 ) \\
 &\ - \frac{f_1 \sqrt{|g|}}{g_{11}g_{22}g_{44}g_{66}} e^3 \w e^5 + \frac{f_2 \sqrt{|g|}}{g_{22}g_{33}g_{44}g_{55}} e^1 \w e^6  \Bigg) \nn\\
= - g_s^{-1} \frac{\tau_2}{|\tau_2|} & \Bigg(- \tilde{g}^{11} \del_1(e^{-4A}) e^2 \w e^5 + \tilde{g}^{22} \del_2(e^{-4A}) e^1 \w e^5 - \tilde{g}^{55} \del_5(e^{-4A}) e^1 \w e^2 \nn\\
 &\ - \frac{f_1 \tilde{g}_{33}}{\tilde{g}_{11}\tilde{g}_{22}} e^3 \w e^5 + \frac{f_2 \tilde{g}_{66}}{\tilde{g}_{22}\tilde{g}_{55}} e^1 \w e^6  \Bigg) \sqrt{\tilde{g}_{11}\tilde{g}_{22}\tilde{g}_{55}} \ \ ,\label{Fsol2}
\eea
where $\del_a$ appear. We deduce
\beq
\d F_2= g_s^{-1} \frac{\tau_2}{|\tau_2|} \Bigg( \tilde{\Delta}(e^{-4A}) + f_1^2 \frac{\tilde{g}_{33}}{\tilde{g}_{11}\tilde{g}_{22}} + f_2^2 \frac{\tilde{g}_{66}}{\tilde{g}_{22}\tilde{g}_{55}}  \Bigg) \sqrt{\tilde{g}_{11}\tilde{g}_{22}\tilde{g}_{55}} ~ e^1 \w e^2 \w e^5 \ .
\eeq
The (satisfied) calibration condition \eqref{calib} and the convention \eqref{defvolbot} give
\beq
{\rm vol}_{||} = \frac{\tau_2}{|\tau_2|} \sqrt{g_{33} g_{44} g_{66}} ~ e^3 \w e^4 \w e^6 \ ,\ {\rm vol}_{\bot} = - \frac{\tau_2}{|\tau_2|} \sqrt{g_{11}g_{22}g_{55}} ~ e^1 \w e^2 \w e^5 \ ,
\eeq
where the free sign of $\tau_2$ is related to the ordering ambiguity in these subvolume forms. We rewrite
\beq
\d F_2= - g_s^{-1} \Bigg( \tilde{\Delta}(e^{-4A}) + f_1^2 \frac{\tilde{g}_{33}}{\tilde{g}_{11}\tilde{g}_{22}} + f_2^2 \frac{\tilde{g}_{66}}{\tilde{g}_{22}\tilde{g}_{55}}  \Bigg) e^{3A} {\rm vol}_{\bot} \ . \label{BIsol2}
\eeq
The same comments can be made as for \eqref{BI2sol1}; in particular \eqref{BIsol2} is in agreement with the general BI \eqref{BI}.\\

On this manifold, we made two more attempts to get solutions. For both, satisfying the SUSY conditions still allows an additional $H$-flux, but its BI does not hold. The first attempt consists of the above vacuum with a further $O_6$ along $e^1 \w e^4 \w e^5$. The algebra and the structure forms are compatible with the second projection, and the calibration condition is satisfied. A second warp factor should be introduced for this source with appropriate powers; in particular $e^A$ is the product of the two warp factors since both sources are space-time filling. The SUSY conditions are then satisfied up to the same constraint \eqref{constraintsol2}. The only remaining condition is the BI, which fails to be satisfied: $\d F_2$ contains new terms, indicating further sources along unappropriate directions. These terms are only canceled at the cost of setting the second warp factor to a constant (partial smearing does not work), erasing any explicit appearance of the second source.

In the second attempt, we start with the above vacuum but smear for simplicity the $O_6$: its warp factor is set to a constant. The $O_6$ appears in any case through the constant shift terms in the BI \eqref{BIsol2}. Motivated by \cite{Andriot:2015aza}, we rather consider a localized $O_4$, along $e^2$: using the previous smeared setting is then nothing but a standard search for solution with $O_4$ on this manifold, but starting with the same SU(2) structure (and changing the dilaton accordingly to the source). Once more, all conditions but the BI can be satisfied, with the constraint \eqref{constraintsol2}. The non-zero RR fluxes are $F_4$ and $F_2$. $\d F_4$ provides a Laplacian leading to a correct BI, but $\d F_2$ suffers from the same problem as in the first attempt: canceling undesired terms sets the $O_4$ warp factor to a constant, making the source effectively disappear. Getting rid of undesired terms in the BI is the most difficult step in the resolution \cite{Grana:2006kf}. These terms are either due to intersecting sources, or to the non-trivial fibrations in solvmanifolds, in which case they are related to the constant shift terms in \eqref{BI2sol1} and \eqref{BIsol2}. We will now comment more on the latter.

\subsection{Subspaces wrapped by the sources and consequences on vacua}\label{sec:propsources}

The origin of the constant shift terms in the RR BI \eqref{BI2sol1} and \eqref{BIsol2}, discussed below \eqref{BI2sol1}, is simple to trace back. Related terms already appear in the fluxes \eqref{Fsol1} and \eqref{Fsol2}. The latter being read from the last SUSY condition of \eqref{SUSY}, the constant terms should be found in $(\d -H\w)(e^{3A-\phi}\im(\Phi_2))$. These terms also appear as the smeared limit (or for simplicity $A=0$) in $\d F$ or the flux. In addition, $H=0$ in our vacua. The constant terms thus originate from $\d \im(\Phi_2)|_{A=0}$, then non-zero. More precisely, they come from the $(p-3)$-form in it, for a flux sourced by an $O_p$. The calibration condition \eqref{calib}, satisfied in the vacua, relates this $(p-3)$-form to the volume form of the subspace wrapped by the source. Denoting $\widetilde{{\rm vol}}_{||}$ its smeared version (with $A=0$), the constant terms are then almost equivalently traced back to
\beq
\d \widetilde{{\rm vol}}_{||} \neq 0 \ . \label{wrappedsubspace}
\eeq
This is verified in the two above vacua: $\d e^3\neq0$ for the first and $\d (e^3\w e^4 \w e^6)\neq0$ for the second one. As pointed-out it below \eqref{BI2sol1}, the sources are then not wrapping cycles. This is still fine with the calibration: calibrated sources should be placed on {\it generalized} cycles \cite{Koerber:2005qi, Martucci:2005ht, Koerber:2006hh, Koerber:2007hd} in presence of fluxes. Note that the Hodge duals, proportional to the ${\rm vol}_{\bot}$ form thanks to \eqref{defvolbot}, are nevertheless closed for both vacua. This allows to integrate the BI on a space without boundary, making the integral of the Laplacian vanish. The integration then gives a standard tadpole condition, equivalently obtained by smearing: the sources contributions are obtained by $\delta \rightarrow 1$ and should equate to the constant terms in $\d F$.

How generic is the property \eqref{wrappedsubspace} for SUSY Minkowski flux vacua on solvmanifolds? When looking at tables of \cite{Grana:2006kf} summarizing all flux vacua found there, one actually notices that \eqref{wrappedsubspace} always holds! In other words, in all flux vacua of \cite{Grana:2006kf}, orientifolds and $D$-branes wrap subspaces whose volume forms are not closed. We believe that this is not a general property of such vacua, but an artefact of the method used in \cite{Grana:2006kf}. As mentioned in section \ref{sec:soltwistedtori}, this method consists in first looking for smeared solutions ($e^A$ and $e^{\p}$ are constant), and secondly trying to localize them, essentially by multiplying by warp factors. With only an $H$-flux, it is not possible to have a Minkowski vacuum on a compact $\mmm$ with a constant dilaton. To get flux vacua with that method, non-zero RR fluxes are then mandatory, and this at the smeared level already, otherwise there is nothing to multiply with. In practice, smeared RR fluxes are given by the constant terms discussed above. As just argued, this requires $(\d -H\w)( \im(\Phi_2)|_{A=0}) \neq 0$: this does not usually hold thanks to the $H$-flux alone, so one rather needs $\d \im(\Phi_2)|_{A=0} \neq 0$. As explained, this leads most of the time to \eqref{wrappedsubspace}. This discussion gives a reason why the method used in \cite{Grana:2006kf} leads to flux vacua with sources along subspaces with non-closed volume forms, a property exemplified in the two previous vacua.

Can one then find a SUSY Minkowski flux vacuum on a solvmanifold, where sources wrap a subspace such that
\beq
\d \widetilde{{\rm vol}}_{||} = 0 \label{wrappedsubspaceclosed}
\eeq
holds? At the end of section \ref{sec:sol2}, we presented an attempt for a solution with an $O_6$ along $e^1\w e^4\w e^5$, which is closed. A reason why this failed was that $\im(\Phi_2)|_{A=0}$ contained, on top of the closed $\widetilde{{\rm vol}}_{||}$, further terms that were not closed. Those lead to undesired terms in the BI, that spoiled the attempt. To get the vacua of interest, it is then safer to ask for the stronger condition $\d \im(\Phi_2)|_{A=0} = 0$. This requirement has several consequences. First, in the smeared limit with $H=0$, the SUSY conditions \eqref{SUSY} become
\beq
\d \Phi_1= 0 \ , \qquad \d \Phi_2= 0 \ . \label{SUSYgenK}
\eeq
The closure of these two pure spinors characterises in GCG the underlying manifold as being Generalized K\"ahler (see e.g. \cite{Gualtieri:2010fd, Sevrin:2013oca, Terryn:2013kr}). A closed $H$-flux adds a twisting to this. Secondly, up to the $H$-flux, the smeared RR fluxes vanish, meaning that we expect them to be given {\it only} by $\del_m e^{-4A}$ terms, and the BI by the Laplacian $\tilde{\Delta}(e^{-4A})$. This absence of constant terms is in agreement with the opposite discussion above \eqref{wrappedsubspace}. Such a situation implies that the number of $O$-planes and $D$-branes are such that they cancel each other in the smeared or integrated BI; the tadpole vanishes thanks to the sources, without constant terms. The sources effectively disappear in the smeared limit. Since sources contributions cancel each other, while having no flux (with $H=0$) and a constant dilaton, we deduce that nothing contributes to the internal Einstein equation in the smeared limit: it boils down to a vanishing Ricci tensor. This implies that the underlying manifold of such vacua needs to be Ricci flat.

To summarize, we asked for SUSY Minkowski flux vacua on solvmanifolds where sources wrap subspaces such that \eqref{wrappedsubspaceclosed} holds. Without much loss of generality, we characterised such vacua as follows: with $H=0$, the underlying manifold is Generalized K\"ahler and Ricci flat, the RR fluxes are only given by derivatives of warp factor, and $O$-planes and $D$-branes contributions cancel each other in the smeared limit. Solutions on $T^6$ with an $O_3$ obtained in \cite{Giddings:2001yu} offer already a good illustration of such vacua. The localized $O_3$ (and $D_3$) is a point in the internal manifold; $\widetilde{{\rm vol}}_{||}$ is then a constant $0$-form, which is certainly closed. We can then verify the characterisation of the vacua: the underlying $T^6$ is a Calabi-Yau, it is thus Ricci flat but also K\"ahler so Generalized K\"ahler; the RR flux $F_5$ is given purely by derivatives of the warp factor (roughly $*\d(e^{-4A})$). In addition, there are optional $H$-flux and $F_3$, related to each other and closed. When non-zero, these would generate a constant term in the $F_5$ BI, that shifts the sources contributions. We are now interested in solvmanifolds different than the torus: as detailed in section \ref{sec:flattwtor}, only three non-trivial algebras \eqref{s1s2s3} give rise to Ricci flat solvmanifolds. We will then look in section \ref{sec:newsol} for vacua satisfying \eqref{wrappedsubspaceclosed} on those solvmanifolds.\footnote{We will show in section \ref{sec:CY} that the Ricci flat solvmanifolds are in particular K\"ahler, hence Generalized K\"ahler. In \cite{Fino07} is presented another six-dimensional solvmanifold that is not K\"ahler but still Generalized K\"ahler. This recalls that we are asking for more than this last GCG characterisation: on top of the SUSY conditions \eqref{SUSYgenK}, the BI is crucial to conclude on the Ricci flatness.} The three algebras were already mentioned in \cite{Grana:2006kf} when discussing ``fluxless solutions'' (see section \ref{sec:soltwistedtori}): those fluxless vacua could then correspond to the smeared limit of the localized flux vacua obtained here.

\section{Truly new vacua on solvmanifolds}\label{sec:newsol}

In this section, we first present new SUSY Minkowski flux vacua on solvmanifolds corresponding to $s_3$ in \eqref{s1s2s3}. The subspaces wrapped by the sources verify the condition \eqref{wrappedsubspaceclosed} presented in details in section \ref{sec:propsources}; this is responsible for the very special form of these vacua and the fact they are localized. Being Ricci flat, solvmanifolds corresponding to $s_1$ and $s_2$ are also candidates for such new vacua, but we did not find any on those as discussed in section \ref{sec:s1s2}. Finally, we show in section \ref{sec:Td} that the new vacua are not T-dual to geometric ones on $T^6$, making them ``truly new'' as defined in section \ref{sec:soltwistedtori}.

As in section \ref{sec:solnil}, the vacua presented here are solutions to all required constraints summarized at the end of section \ref{sec:susysol} and detailed in appendix \ref{ap:condSUSYvac}. These solutions have either SU(3) structure or orthogonal SU(2) structure. In addition, they can be grouped in two families, where solutions within one family should be (Buscher) T-dual to each other, at least for some values of the parameters; we do not verify this explicitly though. The two families differ essentially in the (1,0)-forms parameterizing the structure forms \eqref{SU2forms} and \eqref{SU3forms}, corresponding to different almost complex structures. The solutions are summarized in \eqref{figuresols3}. Our goal here is to provide one example of solution for each $O_p$, and discuss the interesting particularities appearing for each of them. We do not aim at an exhaustive search of solutions: on the contrary, we indicate in several places possibilities for more solutions.

\subsection{With $O_4$}

The only globally defined closed one-forms, as needed for \eqref{wrappedsubspaceclosed}, are $e^5$ and $e^6$. Compatibility of the $O_4$ projection with the algebra excludes $e^6$, so we consider an $O_4$ along $e^5$. $A$ is taken to depend on all coordinates except $y^5$, and $e^{\phi}=g_s e^A$. An $O_4$ cannot admit an SU(3) or intermediate SU(2) structure solution, so we present here one with orthogonal SU(2) structure. The structure forms are given in \eqref{SU2forms} with
\beq
z^3= e^{-A} e^6 + \i e^A \tau_5 e^5 \ ,\ z^1=e^{-A} e^1 + \i e^{-A} \tau_3 e^3\ ,\ z^2=e^{-A} e^2 + \i e^{-A} \tau_4 e^4 \ ,\label{zs3O4}
\eeq
and the metric is
\beq
g_{ab}=\mbox{diag}(e^{-2A} t_1, e^{-2A} t_2, e^{-2A} t_1 (\tau_3)^2, e^{-2A} t_2 (\tau_4)^2, e^{2A} t_3 (\tau_5)^2, e^{-2A} t_3 ) \ .\label{metricsols3O4}
\eeq
The SUSY conditions are satisfied provided
\beq
\tau_4=\tau_3 ~ r\ ,\quad r=\pm 1 \ ,\label{cond1sols3O4}
\eeq
while the BI of $F_4$ will impose further
\beq
t_1=t_2 \ .\label{cond2sols3O4}
\eeq
These two constraints make the smeared metric $\tilde{g}_{ab}$ Ricci flat: as discussed in section \ref{sec:flattwtor}, this is realised with pairs of equal radii $\tilde{g}_{11}=\tilde{g}_{22}$ and $\tilde{g}_{33}=\tilde{g}_{44}$, obtained here with \eqref{cond1sols3O4} and \eqref{cond2sols3O4}. With these constraints, we can still have optional fluxes
\bea
H &= \frac{h}{2} e^{2A} (z^1 \w \ov{z}^2+ \ov{z}^1 \w z^2) \w e^6 = h (e^1\w e^2 + \tau_3 \tau_4 ~ e^3 \w e^4) \w e^6 \\
g_s F_2 &= - \frac{r h}{\sqrt{\tilde{g}_{66}}} \frac{\tau_5}{|\tau_5|} (e^1\w e^2 + \tau_3 \tau_4 ~ e^3 \w e^4) \ ,
\eea
where the constant $h$ can be chosen to vanish. Both fluxes are closed. We have as well
\bea
g_s F_4 = - \frac{\tau_5 \sqrt{|\tilde{g}|}}{|\tau_5| \sqrt{\tilde{g}_{55}}} &\Bigg( - \tilde{g}^{11} \del_1 (e^{-4A}) e^2\w e^3\w e^4\w e^6  + \tilde{g}^{22} \del_2 (e^{-4A}) e^1\w e^3\w e^4\w e^6  \nn\\
&\ - \tilde{g}^{33} \del_3 (e^{-4A}) e^1\w e^2\w e^4\w e^6 + \tilde{g}^{44} \del_4 (e^{-4A}) e^1\w e^2\w e^3\w e^6 \nn\\
&\ - \tilde{g}^{66} \del_6 (e^{-4A}) e^1\w e^2\w e^3\w e^4 \Bigg) \ , \label{Fsols3O4}
\eea
with derivatives $\del_a$ in flat indices, as discussed below \eqref{algs3} or \eqref{F40sol1}. This leads to\footnote{\label{foot:subtle}Applying $\d =e^a\w \del_a$ on $F_4$, one gets the Laplacian discussed in section \ref{sec:twtor}, and two other types of terms. The first ones come from applying $\d$ on the $e^a$ in $F_4$. The second ones appear when applying $e^5 \w \del_5$ of $\d$ on the derivatives in $F_4$: those are in flat indices and thus depend on $y^5$, even if $A$ does not. This can be computed using the algebra \eqref{algs3} to commute $\del_5$ with other $\del_a$. This second set of terms then cancel the first ones, provided the equality of radii mentioned below \eqref{cond2sols3O4} holds, leaving in the end only the Laplacian. Another way to proceed is to work all along in curved (coordinate) indices, starting with $F_4$ and using \eqref{vielbeins3}. The requirement of radii equalities appears again; it allows to obtain the same Laplacian.}
\beq
g_s \d F_4 = \frac{\tau_5 }{|\tau_5|} ~ \tilde{\Delta} (e^{-4A}) ~ \frac{\sqrt{|\tilde{g}|}}{\sqrt{\tilde{g}_{55}}} ~ e^1 \w e^2\w e^3\w e^4\w e^6 \ .
\eeq
The satisfied calibration condition \eqref{calib} and the convention \eqref{defvolbot} give
\beq
{\rm vol}_{||} = \frac{\tau_5}{|\tau_5|} \sqrt{g_{55}} ~ e^5 \ ,\ {\rm vol}_{\bot} = - \frac{\tau_5}{|\tau_5|} \frac{\sqrt{|g|}}{\sqrt{g_{55}}} ~ e^1 \w e^2 \w e^3 \w e^4 \w e^6 \ .
\eeq
The $F_4$ BI is then written
\beq
\d F_4 - H\w F_2 = - g_s^{-1} \left( \tilde{\Delta} (e^{-4A}) + \frac{2 h^2}{\tilde{g}_{11}\tilde{g}_{22}\tilde{g}_{66}}  \right) e^{5A} {\rm vol}_{\bot} \ .\label{BIsols3O4}
\eeq
For $H= F_2 = 0$, i.e. $h=0$, the solution is as expected in section \ref{sec:propsources}: the smeared metric is Ricci flat, $F_4$ only depends on derivatives of the warp factor, it vanishes in the smeared limit so the sources ($O_4$ and $D_4$) cancel each others contribution. Having $h\neq0$ is an additional freedom: it shifts the numbers of $O_4$ and $D_4$ as can be seen in the BI. This acts precisely as the constant terms discussed below \eqref{BI2sol1} and in section \ref{sec:propsources} for the vacua on nilmanifolds.

\subsection{With $O_5$}

We consider an $O_5$ along $e^5\w e^6$, which is closed. Other directions may as well be suited: $(e^1+\tau e^3) \w (e^2+r\tau e^4)$ for a constant $\tau$ and $r^2=1$ is also closed, while solutions with smeared $O_5$ were obtained in \cite{Grana:2006kf, Andriot:2008va}. We only give here one solution with the desired form; it is very probably T-dual to the previous $O_4$ one, upon relating parameters. Our $A$ here does not depend on $y^5, y^6$. The solution has an SU(3) structure. The structure forms are given in \eqref{SU3forms} with
\beq
z^1= e^{-A} (e^1 + \i \tau_3 e^3) \ ,\ z^2= e^{-A} (e^2 + \i \tau_4 e^4) \ ,\ z^3= e^{A} (e^6 + \i \tau_5 e^5) \ ,
\eeq
the corresponding metric being
\beq
g_{ab}=\mbox{diag}(e^{-2A} t_1, e^{-2A} t_2, e^{-2A} t_1 (\tau_3)^2, e^{-2A} t_2 (\tau_4)^2, e^{2A} t_3 (\tau_5)^2, e^{2A} t_3 ) \ .\label{metricsols3O5}
\eeq
The SUSY conditions are satisfied provided
\beq
\tau_4 = r ~\tau_3,\ r^2=1 \ ,
\eeq
while the $F_3$ BI will then require
\beq
t_1=t_2 \ .
\eeq
Related comments were made for the $O_4$ solution. The only non-zero flux is
\bea
g_s F_3 = \frac{\tau_5 \sqrt{|\tilde{g}|}}{|\tau_5| \sqrt{\tilde{g}_{55} \tilde{g}_{66}}} &\Bigg( \tilde{g}^{11} \del_1 (e^{-4A}) e^2\w e^3\w e^4 - \tilde{g}^{22} \del_2 (e^{-4A}) e^1\w e^3\w e^4  \nn\\
&\ + \tilde{g}^{33} \del_3 (e^{-4A}) e^1\w e^2\w e^4 - \tilde{g}^{44} \del_4 (e^{-4A}) e^1\w e^2\w e^3 \Bigg) \ , \label{Fsols3O5}
\eea
expressed as previously with flat $\del_a$, giving
\beq
g_s \d F_3 = \frac{\tau_5 \sqrt{|\tilde{g}|}}{|\tau_5| \sqrt{\tilde{g}_{55} \tilde{g}_{66}}}~ \tilde{\Delta} (e^{-4A})~ e^1 \w e^2\w e^3\w e^4 \ .
\eeq
As above, we verify the calibration condition and obtain
\beq
{\rm vol}_{||}=  \frac{\tau_5 }{|\tau_5| } \sqrt{g_{55} g_{66}} ~ e^5 \w e^6 \ ,\ {\rm vol}_{\bot}= - \frac{\tau_5 \sqrt{|g|}}{|\tau_5| \sqrt{g_{55} g_{66}}} ~ e^1 \w e^2\w e^3\w e^4 \ ,
\eeq
leading to the expected BI
\beq
g_s \d F_3 = - \tilde{\Delta} (e^{-4A})~ e^{4A} {\rm vol}_{\bot}\ .
\eeq

\subsection{With $O_6$}\label{sec:newsolO6}

We present here two vacua with $O_6$. The first one reproduces with a relabeling and slight simplification the solution obtained in section 2.3.1 of \cite{Andriot:2010ju} and mentioned in section \ref{sec:soltwistedtori}. It has an SU(3) structure. The second vacuum is new and has an orthogonal SU(2) structure. Both consider an $O_6$ along $e^1 \w e^2 \w e^5$, which is closed, $A$ depends only on directions $3,4,6$, and $e^{\phi}=g_s e^{3A}$.

The structure forms of our first solution are given in \eqref{SU3forms} with
\beq
z^1=e^{A} e^1 + \i e^{-A} \tau_3 e^3\ ,\ z^2=e^{A} e^2 + \i e^{-A} \tau_4 e^4 \ ,\ z^3= e^{A} e^5 + \i e^{-A} \tau_6 e^6 \ .\label{1formsO6SU3}
\eeq
We deduce the metric
\beq
g_{ab}=\mbox{diag}(e^{2A} t_1, e^{2A} t_2 , e^{-2A} t_1 (\tau_3)^2, e^{-2A} t_2 (\tau_4)^2, e^{2A} t_3, e^{-2A} t_3  (\tau_6)^2) \ .\label{metricsols3O6SU3}
\eeq
The SUSY conditions are satisfied provided
\beq
t_1\tau_3=t_2\tau_4 ~ r\ ,\quad r^2=1 \ .\label{cond0sols3O6SU3}
\eeq
The only non-zero flux is then
\bea
g_s F_2 = \sqrt{\tilde{g}_{33}\tilde{g}_{44}\tilde{g}_{66}} &\Bigg( - \tilde{g}^{33} \del_3 (e^{-4A}) e^4\w e^6  + \tilde{g}^{44} \del_4 (e^{-4A}) e^3\w e^6 - \tilde{g}^{66} \del_6 (e^{-4A}) e^3\w e^4 \Bigg) \nn\\
+ \frac{q \tau_6 \tau_4}{\sqrt{\tilde{g}_{33}\tilde{g}_{44}\tilde{g}_{66}}} & \left(1- \frac{\tau_3~r}{\tau_4} \right) e^{-4A} \left(\tilde{g}_{11}\tilde{g}_{33} e^1 \w e^3 - r \tilde{g}_{22}\tilde{g}_{44} e^2 \w e^4 \right) \label{F2solO6solv}
\eea
with derivatives in flat indices as previously. Computing its exterior derivative and studying the resulting source terms, we find two solutions allowing a suitable BI. The first one sets $A$ to a constant, leading to smeared intersecting sources along $e^1 \w e^4 \w e^6$ and $e^2 \w e^3 \w e^6$: such solutions were already obtained in \cite{Grana:2006kf, Andriot:2008va}. The second option, on which we focus here, is to consider the refined constraints
\beq
\tau_4= \tau_3 ~ r\ ,\quad t_1=t_2\ ,\quad r^2=1 \ .
\eeq
Those make the second row in \eqref{F2solO6solv} vanish, and imply (once again) the Ricci flatness of the smeared metric. In particular, one has $\tilde{g}_{33}=\tilde{g}_{44}$, required to get rid of undesired terms in the BI. We finally obtain
\beq
g_s \d F_2 = - \tilde{\Delta} (e^{-4A}) ~ \sqrt{\tilde{g}_{33}\tilde{g}_{44}\tilde{g}_{66}} ~ e^3 \w e^4\w e^6 \ .
\eeq
The satisfied calibration condition \eqref{calib} and the convention \eqref{defvolbot} give
\beq
{\rm vol}_{||} = \sqrt{g_{11}g_{22}g_{55}} ~ e^1 \w e^2 \w e^5 \ ,\ {\rm vol}_{\bot} =  \sqrt{g_{33}g_{44}g_{66}} ~ e^3 \w e^4 \w e^6 \ .
\eeq
The $F_2$ BI is then written
\beq
\d F_2 = - g_s^{-1} ~ \tilde{\Delta} (e^{-4A}) ~ e^{3A} {\rm vol}_{\bot} \ .\label{BIsols3O6SU3}
\eeq
As a final remark, note that this solution is very probably T-dual to the one with $O_4$, up to relating the parameters. In particular, the (1,0)-forms are very close, with a notable difference in the phase of $z^3$. We believe the latter to be due to a choice of $\theta_-$. This phase is not fixed by the $O_6$ projection conditions and we chose it as indicated in appendix \ref{ap:condSUSYvac} to be $\theta_-=\pi$. Another value could accommodate better the T-duality match.

We now present the vacuum with orthogonal SU(2) structure. Its existence suggests another T-dual solution with $O_4$, or at least another possible almost complex structure there. The structure forms \eqref{SU2forms} are given by
\beq
z^1=e^{A} ( e^1 + \i \tau_2 e^2)\ ,\ z^2=e^{-A} ( e^3 + \i \tau_4 e^4) \ ,\ z^3= e^{A} e^5 + \i e^{-A} \tau_6 e^6 \ .
\eeq
We deduce the metric
\beq
g_{ab}=\mbox{diag}(e^{2A} t_1, e^{2A} t_1 (\tau_2)^2, e^{-2A} t_2, e^{-2A} t_2 (\tau_4)^2, e^{2A} t_3, e^{-2A} t_3  (\tau_6)^2) \ .\label{metricsols3O6}
\eeq
The SUSY conditions are satisfied provided
\beq
\tau_4=-\tau_2 ~ r\ ,\quad r^2=(\tau_2)^2=(\tau_4)^2=1 \ .\label{cond1sols3O6}
\eeq
These constraints make the smeared metric Ricci flat as discussed above (one has $\tilde{g}_{11}=\tilde{g}_{22}$ and $\tilde{g}_{33}=\tilde{g}_{44}$). On the contrary to the previous solutions, this is obtained entirely from the SUSY. More precisely, the BI will also require such equalities, but those are already guaranteed in \eqref{cond1sols3O6}. Its BI forces the $H$-flux to vanish, leaving the only non-zero flux to be
\bea
g_s F_2 = - \frac{\tau_2 \sqrt{|\tilde{g}|}}{|\tau_2| \sqrt{\tilde{g}_{11}\tilde{g}_{22}\tilde{g}_{55}}} &\Bigg( - \tilde{g}^{33} \del_3 (e^{-4A}) e^4\w e^6  + \tilde{g}^{44} \del_4 (e^{-4A}) e^3\w e^6 - \tilde{g}^{66} \del_6 (e^{-4A}) e^3\w e^4 \Bigg) \ ,\nn
\eea
with derivatives in flat indices. This leads to
\beq
g_s \d F_2 = \frac{\tau_2 }{|\tau_2|} ~ \tilde{\Delta} (e^{-4A}) ~ \frac{\sqrt{|\tilde{g}|}}{\sqrt{\tilde{g}_{11}\tilde{g}_{22}\tilde{g}_{55}}} ~ e^3 \w e^4\w e^6 \ .
\eeq
The satisfied calibration condition \eqref{calib} and the convention \eqref{defvolbot} give
\beq
{\rm vol}_{||} = -\frac{\tau_2}{|\tau_2|} \sqrt{g_{11}g_{22}g_{55}} ~ e^1 \w e^2 \w e^5 \ ,\ {\rm vol}_{\bot} = - \frac{\tau_2}{|\tau_2|} \frac{\sqrt{|g|}}{\sqrt{g_{11}g_{22}g_{55}}} ~ e^3 \w e^4 \w e^6 \ .
\eeq
The $F_2$ BI is then written
\beq
\d F_2 = - g_s^{-1} ~ \tilde{\Delta} (e^{-4A}) ~ e^{3A} {\rm vol}_{\bot} \ .\label{BIsols3O6}
\eeq
The orthogonal SU(2) and SU(3) structure solutions presented here are very close, when looking only at the fields ($g, \phi, F_2$) and fixing some parameters. This suggests the existence of intermediate SU(2) structure solutions. It also indicates a possible vacuum with ${\cal N} > 1$ SUSY preserved: such situations were discussed in \cite{Grana:2006kf}.

\subsection{With $O_7$}

We consider an $O_7$ along $e^1 \w e^2 \w e^5 \w e^6$, which is closed. Up to relabeling, this is the only set of directions for which the algebra is compatible with the $O_7$ projection. $A$ only depends on directions $3,4$, and we choose $e^{\phi}=g_s e^{4A}$. The solution to be presented has an SU(3) structure. It is very likely to be T-dual to the previous $O_6$ one with orthogonal SU(2) structure (see the (1,0)-forms). The SU(3) structure forms are given in \eqref{SU3forms} with
\beq
z^1=e^{A} ( e^1 + \i \tau_2 e^2)\ ,\ z^2=e^{-A} ( e^3 + \i \tau_4 e^4) \ ,\ z^3= e^{A} (e^5 + \i \tau_6 e^6) \ .
\eeq
We deduce the metric
\beq
g_{ab}=\mbox{diag}(e^{2A} t_1, e^{2A} t_1 (\tau_2)^2, e^{-2A} t_2, e^{-2A} t_2 (\tau_4)^2, e^{2A} t_3, e^{2A} t_3  (\tau_6)^2) \ .\label{metricsols3O7}
\eeq
The SUSY conditions are satisfied provided
\beq
\tau_4=-\tau_2 ~ r\ ,\quad r^2=(\tau_2)^2=(\tau_4)^2=1 \ ,\label{cond1sols3O7}
\eeq
as in \eqref{cond1sols3O6}. As mentioned there, these constraints give the expected Ricci flatness of the smeared metric. We take $H=0$, implying $F_3=0$, leaving the only non-zero flux to be
\beq
g_s F_1 = \frac{\tau_2 \tau_6 }{|\tau_2 \tau_6| } \sqrt{\tilde{g}_{33}\tilde{g}_{44}} \Bigg( \tilde{g}^{33} \del_3 (e^{-4A}) e^4 - \tilde{g}^{44} \del_4 (e^{-4A}) e^3 \Bigg) \ ,\nn
\eeq
with derivatives in flat indices as previously. Using $ \tilde{g}_{33} = \tilde{g}_{44} $, this leads to
\beq
g_s \d F_1 = \frac{\tau_2 \tau_6 }{|\tau_2 \tau_6| } \sqrt{\tilde{g}_{33}\tilde{g}_{44}} ~ \tilde{\Delta} (e^{-4A}) ~ e^3 \w e^4 \ .
\eeq
The satisfied calibration condition \eqref{calib} and the convention \eqref{defvolbot} give
\beq
{\rm vol}_{||} = -\frac{\tau_2 \tau_6 }{|\tau_2 \tau_6| } \sqrt{g_{11}g_{22}g_{55}g_{66}} ~ e^1 \w e^2 \w e^5 \w e^6 \ ,\ {\rm vol}_{\bot} = - \frac{\tau_2 \tau_6 }{|\tau_2 \tau_6| } \sqrt{g_{33}g_{44}} ~ e^3 \w e^4\ .
\eeq
The $F_1$ BI is then written
\beq
\d F_1 = - g_s^{-1} ~ \tilde{\Delta} (e^{-4A}) ~ e^{2A} {\rm vol}_{\bot} \ .\label{BIsols3O7}
\eeq
Finally, let us point-out that a fluxless solution with $O_7$ is mentioned in \cite{Grana:2006kf}: the structure forms given there differ from the present ones, possibly suggesting further flux solutions.

More generally, the SUSY conditions \eqref{SUSYSU3O7} leave room for non-zero $H$ and $F_3$. One consequence in the BI would be the non-trivial $H\w F_3$ contribution to that of $F_5$: it would indicate smeared $O_3$ and $D_3$. Having a simultaneous $O_3$ is a priori possible: the projection conditions are known to be the same as those of the $O_7$ \eqref{O7proj}. However, an $O_3$ projection is not compatible with the algebra. This excludes having an $H$-flux here. An analogous extension by an $H$-flux will nevertheless be realised for the next vacuum, with $O_8$ and $O_4$.

\subsection{With $O_8$}

We present here a solution with $O_8$, very likely to be T-dual to the above one with $O_4$, given the set of (1,0)-forms. Another vacuum, T-dual to the $O_6$ one with orthogonal SU(2) structure, may as well exist. The only set of directions for which the $O_8$ projection is compatible with the algebra is $e^1 \w e^2 \w e^3 \w e^4 \w e^5$, which is also closed. Considering the $O_8$ along it, we take $A(y^6)$, as well as $e^{\phi}= g_s e^{5A}$. As for $O_4$, only orthogonal SU(2) structures are allowed \cite{Grana:2006kf}. The structure forms are given by \eqref{SU2forms} with
\beq
z^3= e^{-A} e^6 + \i e^A \tau_5 e^5 \ ,\ z^1=e^{A} (e^1 + \i \tau_3 e^3)\ ,\ z^2=e^{A} (e^2 + \i \tau_4 e^4) \ .\label{zs3O8}
\eeq
We deduce the metric
\beq
g_{ab}=\mbox{diag}(e^{2A} t_1, e^{2A} t_2, e^{2A} t_1 (\tau_3)^2, e^{2A} t_2 (\tau_4)^2, e^{2A} t_3 (\tau_5)^2, e^{-2A} t_3 ) \ .\label{metricsols3O8}
\eeq
The SUSY conditions are satisfied provided
\beq
\tau_4=\tau_3 ~ r\ ,\quad r=\pm 1 \ .\label{cond1sols3O8}
\eeq
As for the $O_4$ vacuum, some fluxes are optional; we will come back to them. One constraint on the $H$-flux is $H\w j=0$, leaving in any case only one contribution to $F_0$ \eqref{SUSYSU2}: we obtain
\beq
g_s F_0 = r \frac{\tau_5 }{|\tau_5|} \sqrt{\tilde{g}_{66}} ~ \tilde{g}^{66} \del_6 (e^{-4A}) \ , \label{Fsols3O8}
\eeq
giving
\beq
g_s \d F_0 = r \frac{\tau_5 }{|\tau_5|} ~ \tilde{\Delta} (e^{-4A}) ~ \sqrt{\tilde{g}_{66}} ~ e^6 \ .
\eeq
The satisfied calibration condition \eqref{calib} and the convention \eqref{defvolbot} give
\beq
{\rm vol}_{||} = r \frac{\tau_5}{|\tau_5|} \sqrt{g_{11}g_{22}g_{33}g_{44}g_{55}} ~ e^1 \w e^2 \w e^3 \w e^4 \w e^5 \ ,\ {\rm vol}_{\bot} = - r\frac{\tau_5}{|\tau_5|} \sqrt{g_{66}} ~ e^6 \ .
\eeq
The $F_0$ BI is then written
\beq
\d F_0  = - g_s^{-1} \tilde{\Delta} (e^{-4A}) ~ e^{A} {\rm vol}_{\bot} \ .\label{BIsols3O8}
\eeq
Possible additional fluxes are $H$ and $F_2$. Their BI and the SUSY conditions provide various constraints. In particular, the $F_2$ BI, independently of $H$, imposes
\beq
t_1=t_2 \ ,\label{cond2sols3O8}
\eeq
giving once again the expected Ricci flatness. The freedom in the fluxes is in the end restricted to a constant $h$ with
\bea
H &= \frac{h}{2} e^{6A} (z^1 \w \ov{z}^2+ \ov{z}^1 \w z^2) \w e^6 = h e^{8A} (e^1\w e^2 + \tau_3 \tau_4 ~ e^3 \w e^4) \w e^6 \\
g_s F_2 &= - \frac{r h}{\sqrt{\tilde{g}_{66}}} \frac{\tau_5}{|\tau_5|} e^{4A} (e^1\w e^2 + \tau_3 \tau_4 ~ e^3 \w e^4) \ .
\eea
Both fluxes satisfy their BI
\beq
\d H=0 \ , \quad \d F_2 - H F_0 = 0 \ ,
\eeq
with the above $F_0$ \eqref{Fsols3O8}. Finally, these fluxes contribute as well to the $F_4$ BI through
\beq
\d F_4 - H\w F_2 \ .
\eeq
SUSY gives here $F_4=0$, but $H\w F_2 \neq 0$ (for $h\neq0$) and is given by the five-form orthogonal to $e^5$. So this term would indicate smeared $O_4$ and $D_4$ along $e^5$. The previous $O_4$ vacuum actually considered sources along $5$. In addition, the (1,0)-forms and structure forms of both solutions are the same up to warp factors. The projection conditions for an $O_8$ and an $O_4$ are also the same, as given in \eqref{O4proj}. We conclude that our $O_8$ solution is perfectly compatible with having additional smeared $O_4$ and $D_4$ along $e^5$; as usual, a non-zero $h$ would shift their number. It would be interesting to look for a localized version of this vacuum with the two kinds of sources.

\subsection{No vacuum for $s_1$ and $s_2$}\label{sec:s1s2}

We looked for vacua of the kind presented at the end of section \ref{sec:propsources}, with SU(3) or orthogonal SU(2) structure, on solvmanifolds corresponding to $s_1$ and $s_2$ in \eqref{s1s2s3}, and did not find any. The more mathematical results presented in section \ref{sec:CY}, in particular those of \cite{Fino} on the absence of solution to $\d \Omega=0$, confirm this negative conclusion. We summarize here our attempts.

$s_2$ is constraining: no $O_4$, $O_6$, or $O_8$ is compatible with the related algebra. An $O_5$ is compatible with several sets of directions, but the only closed one is $e^3 \w e^6$. The same holds for $O_7$ with $e^1 \w e^2 \w e^3 \w e^6$, up to relabeling. In both settings, we did not find an SU(3) or orthogonal SU(2) structure solution. The absence of $O_5$ solution is claimed in \cite{Grana:2006kf}, but the existence of a fluxless vacuum with $O_7$ along $e^1 \w e^2 \w e^3 \w e^6$ and an SU(3) structure is mentioned there; that solution is nevertheless not given.

$s_1$ is a priori less constraining for the orientifold projection, but SUSY conditions still prevent us from finding solutions, as we briefly describe. Consider $e^1$ and $e^2$: they enter only three closed two-forms, namely $e^1\w e^2$, $e^1\w e^3$, $e^2\w e^3$. $\omega$ of an orthogonal SU(2) structure depends at least on four independent real two-forms. In addition, in type IIA, the smeared limit implies $\d \omega=0$, meaning that these four two-forms are closed. Given the form of $\omega$ (built from one-forms), we deduce that it cannot depend on $e^1$ and $e^2$. These one-forms must then both be in $z$. They are however not closed, which makes it impossible to satisfy in the smeared limit $\d (\re (z))=0$. Type IIA solutions with orthogonal SU(2) structure are then excluded. It is certainly as difficult to satisfy for an SU(3) structure $\d \Omega=0$. That equation appears in the smeared limit in type IIB, but also for our solutions with $O_6$ for which we expect $F_2$ to vanish in that limit. We are then only left with a possible orthogonal SU(2) structure solution and $O_5$ or $O_7$. The sets of directions compatible with the projection and closed are respectively $e^3 \w e^4$, and $e^1 \w e^2 \w e^3 \w e^6$ or $e^3 \w e^4 \w e^5 \w e^6$, up to relabeling. The corresponding projection conditions nevertheless restrict the structure forms in such a way that the SUSY conditions cannot be satisfied. An analogous comment is made in \cite{Grana:2006kf} on the $O_5, O_6, O_7$ projections.

\subsection{T-dualising the vacua and (no) relation to the torus}\label{sec:Td}

As discussed at the end of section \ref{sec:soltwistedtori}, it is important to determine whether vacua obtained are related by T-duality to other (geometric) vacua on a torus. If not, they would lead after dimensional reduction to new physics, and are thus called ``truly new'' vacua. As mentioned for our vacua on nilmanifolds and first shown in \cite{Kachru:2002sk}, a (Buscher) T-duality can remove a non-trivial fibration, e.g. transform a nilmanifold into a torus, and bring the information into a $b$-field. For nilmanifolds, the off-diagonal metric encoding the fibration is typically linear in a base coordinate. The T-duality brings the linearity to a $b$-field, that is then well-defined as it patches with gauge transformations, leading to a globally defined $H$-flux. From a four-dimensional point of view, the ``geometric flux'' $f^3{}_{12}$ becomes an $H_{123}$, as realised starting with the vacuum on the Heisenberg manifold. As we now show, several aspects of this scenario do not work for solvmanifolds, especially for the vacua obtained previously in section \ref{sec:newsol}; those are then truly new.

In the vacua obtained previously, the $H$-flux, if any, can always be set to zero. We do so here, considering no $b$-field, and we rather focus only on the metric of the solvmanifold. As described for nilmanifolds, the T-dual metric could be that of a torus; having an initial non-zero $b$-field cannot help in getting this result. The vacua obtained share the particularity of having a Ricci flat metric in the smeared limit: this translates into pairs of equal radii such as $\tilde{g}_{11}=\tilde{g}_{22}$. It is enough to focus on this pair of directions, and work in the smeared limit: the question is whether the T-duality can remove the non-trivial solvmanifold fibration, which can be studied with a pair of such directions, and is independent of the warp factor. The radii equality, combined with the vielbeins being just local rotations \eqref{vielbeins3}, gives a surprising metric along these two directions
\beq
\d \tilde{s}^2_{12}= \tilde{g}_{11} (e^1)^2 + \tilde{g}_{22} (e^2)^2 = \tilde{g}_{11} \left( (e^1)^2 + (e^2)^2 \right) = \tilde{g}_{11} \left( (\d y^1)^2 + (\d y^2)^2 \right) \ . \label{funnymetric}
\eeq
The information on the fibration seems to disappear (in particular, there is no off-diagonal component), this metric looks like that of a torus! Naively T-dualising along $y^1$ or $y^2$ or both would then simply result in inverting $\tilde{g}_{11}$ to $1/\tilde{g}_{11}$. Both before and after the T-duality, our vacua on solvmanifolds then seem easily related to the torus with various radii. This is however not correct as we now explain, recalling the discussion on the lattice around \eqref{cohomo}. According to the lattice chosen, in particular the value of $q$, one gets different solvmanifolds. If $q$ is an integer, the manifold is a torus: the $\d y^m$ are globally well-defined, and our vacua are simply on a torus to start with; this is not the case we are interested in. Choosing another value for $q$ as specified in \cite{Console:2012wp, Grana:2013ila}, e.g. half integer, etc., the solvmanifold is really different than a torus: $\d y^1$ and $\d y^2$, as well as coordinates $y^1$ and $y^2$, are not globally defined, but experience a monodromy related to the non-trivial fibration. \eqref{funnymetric} looks thus locally identical to a torus metric, but is globally different as the $\d y^m$ are not the same. T-dualising along the two directions inverts both $\tilde{g}_{11}$ while preserving the sum of squares: this allows to reconstruct the $e^a$, that are globally defined. So this T-duality along both directions does not relate to the torus, but leaves the solvmanifold to itself up to a change (inversion) of radius.

An alternative to understand this result is through the generalized vielbein $\eee$, that allows to study the effect of T-duality on the vielbein $e$ instead of the metric. We refer for instance to section 4.1.3 of \cite{Andriot:2014uda} for conventions. The generalized metric $\hhh$ is T-dualised into $\hhh'$ by the O(N,N) T-duality transformation $O$ as $\hhh'=O^T \hhh O$. Writing $\hhh=\eee^T \mathbb{I} \eee$, one deduces the T-dual generalized vielbein $\eee'=K\eee O$, where $K$ is an O(N)$\times$O(N) transformation such that $K^T \mathbb{I} K = \mathbb{I}$. The freedom in $K$ is typically used to build a consistent vielbein after T-duality: see the examples below. We start with the previous vacua without $b$-field: $\eee$ is then made of diagonal blocks $e$ and $e^{-T}$, with $e$ a vielbein associated to the solvmanifold (smeared) metric. As above, we focus only on one pair of directions with radii equality. Introducing $R \equiv \sqrt{\tilde{g}_{11}}$, we have
\beq
e= R \times \begin{pmatrix} \cos (q y^5) & -\sin (q y^5) \\ \sin (q y^5) & \cos (q y^5) \end{pmatrix} \ . \label{vielbeinstart}
\eeq
On the torus, a vielbein would be $e=R \times \id_2$. With an appropriate transformation ($O=\id$ and some $K$), one could imagine bringing the vielbein \eqref{vielbeinstart} to that of the torus. The corresponding $K$ would nevertheless be local and depend on $\cos (q y^5)$. This is where the choice of lattice and value of $q$ matter. If $q$ is an integer, this function is globally defined, and so is $K$. The transformation is then allowed, the vielbeins can be mapped, but the manifold is nothing but a torus to start with. If $q$ takes another value mentioned above, the function and thus $K$ are not globally defined. Then, one cannot identify the vielbeins, nor the manifolds. The same reasoning applies after T-duality. A Buscher T-duality along both directions, together with a required $K=O$ as argued below, simply exchanges $e$ with $e^{-T}$: this results only in the inversion $R \rightarrow \frac{1}{R}$, leaving the rotation matrix present in the vielbein. The manifold after T-duality thus remains the same, up to a radius inversion.

The study of T-duality along a single direction of the pair is more technical, and we leave it to appendix \ref{ap:Td}. A crucial point is again whether $K$ is globally defined as discussed in \cite{Grana:2008yw}, from which we conclude that such a T-duality does not lead to a geometric vacuum on the torus. More generally, we conclude that the vacua on solvmanifolds obtained previously in section \ref{sec:newsol} are not T-dual to geometric vacua on a torus. They are rather T-dual to geometric vacua on the same solvmanifold, as indicated in the solutions summary \eqref{figuresols3}. As a consequence, they are truly new vacua and lead to new physics after compactification.

\section{Calabi-Yau and Ricci flat K\"ahler solvmanifolds}\label{sec:CY}

In this section, we show that solvmanifolds corresponding to $s_3$ \eqref{s1s2s3} are Calabi-Yau, while those corresponding to $s_1$ and $s_2$ are ``only'' Ricci flat K\"ahler; we further comment on these results. Let us first recall some required material. The forms $J$ and $\Omega$ defining an SU(3) structure are globally defined and non-vanishing, and are given in this paper by the expressions \eqref{SU3forms}, repeated here for convenience
\beq
J = \frac{\i}{2} \left( t_1~ z^1 \w \ov{z}^1 + t_2~ z^2 \w \ov{z}^2 + t_3~ z^3 \w \ov{z}^3 \right) \ ,\ \Omega= \sqrt{t_1 t_2 t_3}~ z^1 \w z^2 \w z^3 \ ,\ t_{1,2,3} > 0 \ . \label{SU3formsCY}
\eeq
They depend on $z^i, \ov{z}^i$, which are (1,0) or (0,1)-forms with respect to an almost complex structure. We always construct these one-forms so that they are globally defined, at least in the smeared limit, and form a basis. For the almost complex structure to be integrable, further properties on their derivatives are needed, summarized by the condition $\d \Omega = W_5\w \Omega$ (see e.g. \cite{Grana:2005jc}) with $W_5$ a real one-form. The manifold is then complex, and the (1,1)-form $J$ is the corresponding fundamental or Hermitian form. If in addition $\d J=0$, the Hermitian manifold is a K\"ahler manifold. Finally, $\Omega$ being a (3,0)-form in this six-dimensional manifold, $\d \Omega=0$ is equivalent to $\Omega$ being holomorphic. The conditions
\beq
\d J=0 \ , \qquad \d \Omega=0 \ , \label{CYcond}
\eeq
are thus defining a six-dimensional Calabi-Yau: a K\"ahler manifold admitting a nowhere vanishing holomorphic (3,0)-form.

In the vacua described at the end of section \ref{sec:propsources} and obtained in section \ref{sec:newsol} on solvmanifolds corresponding to $s_3$, we can always consider $H=0$, and take in addition the smeared limit $A=0$: the particularity of our solutions is precisely that this results in $F_k=0$ (on top of a constant dilaton). In such a situation, the SUSY conditions with SU(3) structure for an $O_5$ \eqref{SUSYSU3O5}, $O_6$ \eqref{SUSYSU3O6}, or $O_7$ \eqref{SUSYSU3O7}, become simply \eqref{CYcond}, meaning those of a Calabi-Yau. This implies that our solutions with SU(3) structure and $O_5$, $O_6$, or $O_7$, provide in such limits examples of $J$ and $\Omega$ satisfying \eqref{CYcond}. We read from there in the smeared limit two concrete examples given by
\bea
& z^1= e^1 + \i \tau_3 e^3\ ,\ z^2=e^2 + \i r \tau_3 e^4 \ ,\ z^3= e^5 + \i \tau_6 e^6 \ ,\ t_1=t_2\ , \label{zis3CY}\\
& z^1= e^1 + \i \tau_2 e^2\ ,\ z^2=e^3 - \i r \tau_2 e^4 \ ,\ z^3= e^5 + \i \tau_6 e^6 \ ,\ (\tau_2)^2=1\ , \nn
\eea
with $r^2=1$. We conclude that {\it the solvmanifolds associated to $s_3$ are Calabi-Yau}! As mentioned already, our vacua have precisely a Ricci flat smeared metric, built from the SU(3) structure forms as described around \eqref{SU3forms}. This is expected from a Calabi-Yau: it can have different metrics, but one is Ricci flat and related to the forms in \eqref{CYcond}.

In the final stage of this project, we became aware of overlapping results in the recent math literature, that we now mention. In proposition 2.6 or theorem 2.8 of \cite{Fino} was proven that the algebra related to $s_3$ admits a complex structure, and that there exists a corresponding closed $\Omega$. In addition, proposition 3.3 implies that the two complex structures corresponding to \eqref{zis3CY} should be isomorphic; a related $\Omega$ is given in their table 1. Finally, in remark 4.2 of that paper, the algebra corresponding to $s_3$ is mentioned to admit Calabi-Yau metrics, and the related K\"ahler form is given. Further interesting results can be found in theorem 6.4 of \cite{Fino2}. Note that here, we go beyond the algebra or the group, to the solvmanifolds, and this last step is not trivial as we now recall.

As discussed in sections \ref{sec:twtor} and \ref{sec:flattwtor}, the $e^a$ are first defined, equivalently here, on the cotangent bundle of the solvable algebra or of the group. To reach the (compact) solvmanifold, one should further divide by the lattice. The $e^a$ are invariant under the lattice action, by definition, and are thus globally defined on the resulting solvmanifold. The conditions \eqref{CYcond} to be a Calabi-Yau are then not only satisfied by the algebra and the group, but also by the solvmanifold (the structure descends to the latter). However, as discussed in section \ref{sec:flattwtor}, different lattices, related to the value of $q$, can lead here to inequivalent solvmanifolds: three different topologies were identified, as given by the cohomologies in \eqref{cohomo}. One of them makes the solvmanifold simply a torus; others, of interest here, make it different. We then claim that {\it all solvmanifolds obtained from $s_3$ are Calabi-Yau}: this is trivial for the torus, but not for the other cases.

Let us now characterise more these Calabi-Yau solvmanifolds. Since they contain a free circle along the sixth direction, they are not simply connected. As a consequence, they are not in the Kreuzer-Skarke list: in the latter, only sixteen manifolds are not simply connected, and none of those has a vanishing Euler characteristic \cite{He:2013ofa} as ours (see below \eqref{cohomo}). They are also not complete intersection Calabi-Yau manifolds, as those are simply connected. In addition, our solvmanifolds take the form of a quotient, so they are parallelizable (once again, the $e^a$ are globally defined on the quotient). This implies that their structure group is trivial.\footnote{This should not be confused with the SU(3) or SU(2) structure forms involved in the SUSY vacua: to find a solution to the constraints, these forms have to solve differential conditions, while identifying the structure group is only a topological statement.} By definition, their holonomy group is included in SU(3), but since they are not simply connected, the inclusion is strict, i.e. the holonomy group is reducible. It is then either SU(2) or the identity. Manifolds in the first case are often considered to be given locally by the product of a K3 surface and a flat two-dimensional manifold; our Calabi-Yau solvmanifolds are thus more likely to have a trivial holonomy group. This is confirmed by the fact they admit locally a constant $\mathbb{R}^6$ metric (see \eqref{funnymetric}): the Ricci flat metric is actually flat. These Calabi-Yau solvmanifolds can thus be viewed as $\mathbb{R}^6$ with discrete identifications. In view of the discussion at the end of section \ref{sec:flattwtor}, they also admit a torus as a covering space. They are though not a torus orbifold, since they do not have any fixed point or singularity.\\

Are there further solvmanifolds that are Calabi-Yau? As explained in section \ref{sec:flattwtor}, only three algebras, related to $s_1, s_2, s_3$ \eqref{s1s2s3}, lead to manifolds admitting a Ricci flat metric. As detailed in section \ref{sec:s1s2}, finding solutions to $\d \Omega=0$ looks impossible for $s_1$, and we did not find any either for $s_2$, although additional constraints were considered. This is confirmed by proposition 2.6 or theorem 2.8 of \cite{Fino}: algebras (and thus groups) corresponding to $s_1$ and $s_2$ are shown not to admit a complex structure that allows for a closed $\Omega$. We deduce that these solvable groups are not Calabi-Yau. We do not expect this result to change for the solvmanifolds obtained after dividing by a lattice, with the obvious exception of the torus (for the latter, the Calabi-Yau structure thus does not descend from the groups). An argument in favour of this is that the only two cohomologies of solvmanifolds corresponding to $s_1$, given below \eqref{cohomo}, are either that of the torus or that of the algebra. To conclude, leaving aside the tori from $s_1$ and $s_2$, the solvmanifolds corresponding to $s_3$ are very likely to be the only ones that are Calabi-Yau.

The structures carried by the algebras, groups, and some solvmanifolds associated to $s_1$ and $s_2$, are nevertheless interesting as we now show. Consider the following sets of one-forms building the SU(3) structure forms \eqref{SU3forms}
\bea
{\rm For}\ s_1:&~ z^1= e^1 + \i e^2\ ,\ z^2=e^3 + \i e^4 \ ,\ z^3= e^5 + \i e^6 \ ,\\
{\rm For}\ s_2:&~ z^1= e^1 + \i e^2\ ,\ z^2=e^4 + \i e^5 \ ,\ z^3= e^3 + \i e^6 \ .
\eea
The pair(s) of radii being equal, these forms induce a Ricci flat metric. For instance with $s_1$, one has $\tilde{g}_{11}=\tilde{g}_{22}=t_1$, and the Ricci tensor can be obtained from \eqref{Riccitensors3} setting $r=0$ and replacing the direction $5$ by $3$: it is then clearly vanishing. We now compute the exterior derivatives of structure forms and get
\bea
{\rm For}\ s_1:&~ \d J= 0\ ,\ \d \Omega= \frac{\i}{2} q ~ \ov{z}^2 \w \Omega \ , \label{extders1s2}\\
{\rm For}\ s_2:&~ \d J= 0\ ,\ \d \Omega= \frac{\i-r}{2} q ~ \ov{z}^3 \w \Omega \ .\nn
\eea
This can be rephrased with $\d \Omega= W_5 \w \Omega$ where\footnote{Note that both $W_5$ are not exact, so $\Omega$ is not conformally closed.}
\bea
{\rm For}\ s_1:&~ W_5= \frac{\i}{2}q (\ov{z}^2- z^2)= q e^4 \ ,\label{W5}\\
{\rm For}\ s_2:&~ W_5= \frac{1}{2}q (\i\ov{z}^3- \i z^3 - r (\ov{z}^3+ z^3 ) )= q (e^6 -r e^3) \ .\nn
\eea
Recalling the material detailed at the beginning of this section, we read from \eqref{extders1s2} that both almost complex structures are integrable, and the two manifolds are then K\"ahler; this was mentioned for $s_1$ in example 4 of \cite{Hasegawa}, but is a new result for $s_2$. We also explained that the two induced metrics are Ricci flat, so we have {\it Ricci flat K\"ahler manifolds}. In a sense, we cannot get closer to a Calabi-Yau without being Calabi-Yau; another example of such manifolds is given by the Enriques surfaces. This distinction is only possible because these manifolds are not simply connected: indeed, a simply connected compact K\"ahler manifold with vanishing first real Chern class is a Calabi-Yau.

It is worth noticing that if the only non-zero torsion class of the SU(3) structure is $W_5$, then the manifold, being K\"ahler but not Calabi-Yau, can be Ricci flat, provided some constraints on $W_5$ are satisfied. To illustrate this point, we read in \cite{Bedulli} the Ricci tensor and scalar of the metric induced by an SU(3) structure in terms of the torsion classes;\footnote{Most of the physics papers providing analogous formulas consider half-flat manifolds, that sets $W_5=0$. A formula for the Ricci scalar in terms of SU(3)$\times$SU(3) pure spinors was nevertheless derived as (4.20) in \cite{Cassani:2008rb}, imposing few restrictions as e.g. (4.19) there. A general formula without these restrictions was then obtained as (C.1) of \cite{Lust:2008zd}. This led in the subcase of an SU(3) structure to the expression (3.10) of \cite{Held:2010az}. From the latter, one should be able to reproduce \eqref{RW5}. In their notations, $u\sim W_5 |\Omega|^2$ (this $u$ was set to zero in \cite{Cassani:2008rb}). This allows in (3.10) of \cite{Held:2010az} to cancel $|\d \Omega|^2$ by $u^2$ (up to some normalisation), leaving in ${\cal R}$ only $\nabla^m u_m$. The latter is very likely to match $* \d (* W_5)$. In case of an SU(2) structure, see the recent \cite{Solard:2016mej}.} keeping only $W_5$, the scalar is given by
\beq
{\cal R}= 2 \, \d^* W_5 = - 2 * \d (* W_5) \ ,\label{RW5}
\eeq
where we worked-out the adjoint of the exterior derivative $\d^*$ in the present context. Having $W_5 \neq 0$ still allows for $\d (* W_5)=0$, that then leads to a vanishing Ricci scalar. It is straightforward, using \eqref{Hodge}, to verify this last condition for the two previous examples \eqref{W5}; we recall that we also obtained for those a vanishing Ricci tensor.

\section{Outlook}

The new localized vacua of section \ref{sec:newsol} are not T-dual to a vacuum on the torus. The corresponding four-dimensional physics should then be new as stressed in the Introduction, so performing the dimensional reduction would be very interesting. To that end, one should note the following: first the solvmanifolds are Calabi-Yau, and second they are Lie algebra based manifolds that carry, in the language of gauged supergravity, ``geometric flux'' $f^a{}_{bc}$. Both points provide various tools developed for dimensional reduction. Some subtleties should however be taken into account regarding the gauged supergravity point of view, a first one being related to the lattice choice discussed in section \ref{sec:flattwtor}, that affects the reduction as described in \cite{Grana:2013ila}. In addition, the gauged supergravity description typically captures the smeared aspects of vacua; however the particularity of our vacua is precisely that fluxes are present when non-smeared. Related remarks on the role of the warp factor and discrepancies of the reduction with respect to gauged supergravities were made in \cite{Danielsson:2013qfa}. References therein are also of interest, in particular \cite{Martucci:2009sf, Martucci:2014ska} for contributions of the warp factor and \cite{Blaback:2010sj} on relations between smeared and localized vacua. We hope to come back to this dimensional reduction.

Our vacua sit on a T-duality orbit different than that of geometric vacua on the torus. Studying such orbits and resulting physics, especially moduli stabilization, has recently gained attention in the context of non-geometry, where in particular orbits with only non-geometric vacua are looked for. Our vacua are geometric, but it remains interesting that they sit on a new orbit: they could lead after T-duality to new kinds of non-geometric vacua. We failed though to obtain in appendix \ref{ap:Td}, from a ten-dimensional perspective, a characterisation of the latter in terms of (non)-geometric fluxes, even if the result is clear in four dimensions.

The new vacua presented in this paper should allow for interesting extensions or use, in string theory and phenomenology. First, having new and explicit Calabi-Yau manifolds at hand should find various applications. Having in addition localized sources, our vacua could provide interesting settings to build intersecting brane models. They might as well be useful to find fully localized solutions with intersecting sources, as those of \cite{Assel:2011xz, Assel:2012cj, Rota:2015aoa}. Deforming our vacua may also lead to de Sitter vacua, or help in realising inflation scenarios, in which cases having studied the dimensional reduction as discussed above would be important for stability. Finally, it would be interesting to find more localized vacua not T-dual to a vacuum on the torus: further Lie algebra based manifolds mentioned in section \ref{sec:soltwistedtori}, or toric varieties as those discussed in \cite{Larfors:2013zva} and references therein, are interesting candidates for $\mmm$.\\

Last but not least, let us comment on natural seven-dimensional extensions from $s_3$, and manifolds admitting G$_2$-structures. Consider the following seven-dimensional one-forms and relations
\bea
s_4:&\ \d e^1= q~ e^2 \w e^7 \ ,\ \d e^2= -q~ e^1 \w e^7 \ ,\\
&\ \d e^3= rq~ e^4 \w e^7 \ ,\ \d e^4= - rq~ e^3 \w e^7\ ,\nn\\
&\ \d e^5= sq~ e^6 \w e^7 \ ,\ \d e^6= - sq~ e^5 \w e^7\ ,\nn\\
&\ \d e^7= 0 \ ,\nn
\eea
with both $q, r \in \mathbb{R}^*$. We add to $s_3$ an analogous fibration along two further directions. The related solvable groups admit lattices, that give after quotient (compact) seven-dimensional solvmanifolds. Consider first $s=0$: the solvmanifolds are then a circle times the six-dimensional manifolds obtained from $s_3$, the latter being Calabi-Yau. Some of these seven-dimensional solvmanifolds are a priori different than $T^7$, but still have a reduced holonomy. With the above Ricci flat metric for $s_3$ and $r^2=1$, we then get explicit seven-dimensional compact manifolds admitting a G$_2$-structure. Thanks to the Calabi-Yau structure, this G$_2$-structure is parallel, meaning the three-form is closed and coclosed. To construct explicitly the G$_2$ forms, see e.g. \cite{Kaste:2003zd, FinoG2, Freibert}. Consider now $s \in \mathbb{R}^*$. Taking the same radii along $e^1$ and $e^2$, $e^3$ and $e^4$, $e^5$ and $e^6$, should again provide a Ricci flat metric. It would be interesting to study whether, upon probable restrictions on $s$, one obtains again compact manifolds admitting a parallel G$_2$-structure. Shortly after the present paper, the work \cite{Manero} appeared: there, G$_2$-structures on seven-dimensional solvable algebras are studied, and the G$_2$ forms are constructed from SU(3)-structure ones. That paper confirms the present result that $s_4$ with $s=0$ admits a parallel G$_2$-structure; in addition, the same holds for $s=-1-r$.

\vspace{0.4in}

\subsection*{Acknowledgements}

I would like to thank D. Junghans for valuable discussions related to section \ref{sec:propsources}, as well as M. Larfors, R. Minasian, E. E. Svanes, A. Thorne and D. Tsimpis for helpful exchanges. I also thank D. Cassani, A. Fino, J. Gray, V. Manero, C. Shahbazi, H. Triendl and L. Ugarte for further exchanges helping me to improve the paper into its revised version. In addition, I thank the organisers, lecturers and participants to the inspiring workshop ``Special Geometric Structures in Mathematics and Physics'' that took place in Hamburg, Germany, in September 2014. Finally, I thank the great Jacob-und-Wilhelm-Grimm-Zentrum library in Berlin, Germany, in particular the Forschungslesesaal, where part of this project was realised. This work is part of the Einstein Research Project ``Gravitation and High Energy Physics'', which is funded by the Einstein Foundation Berlin.

\newpage

\begin{appendix}

\section{Getting the vacua of interest}\label{ap:condSUSYvac}

We presented in section \ref{sec:susysol} vacua of type II supergravities we are interested in, namely SUSY vacua with fluxes on Minkowski times a compact manifold $\mmm$. We gave the conditions to be verified to get these vacua, as summarized in the list at the end of that section. We provide here pragmatic details for some of these conditions.

\subsection*{Orientifold projection and structure forms}

The orientifold involution $\sigma$ transforms the internal spinors or equivalently $\Phi_{\pm}$ as given in \cite{Grana:2006kf}. According to which pair of pure spinors among \eqref{purespinorstruct} and which $O_p$, we get different transformations of the structure forms and some phases get fixed. We give here the constraints to be verified for some of these cases, that are relevant for this paper. For those, $\theta_-$ is not fixed by the conditions and we take it to be $\theta_-=\pi$. For an SU(2)$_{\bot}$ structure, we get
\bea
O_4/O_8 :&\quad \sigma(\re(z))= - \re(z),\ \sigma(\im(z))= \im(z),\ \sigma(\omega)= \omega, \ \sigma(j)=j \ ,\label{O4proj}\\
O_6:&\quad \sigma(\re(z))= \re(z),\ \sigma(\im(z))= -\im(z),\ \sigma(\omega)= -\omega, \ \sigma(j)=j \ .\label{O6proj}
\eea
For an SU(3) structure in type IIB, the $O_5$ projection imposes $e^{\i \theta_+}$ to be real and we choose $\theta_+=0$ while the $O_7$ imposes it to be purely imaginary and we take $\theta_+=\frac{\pi}{2}$. We get furthermore
\bea
O_5 :& \quad \sigma(J)=J,\ \sigma(\Omega)=\Omega \label{O5proj} \ ,\\
O_6 :& \quad \sigma(J)=-J,\ \sigma(\Omega)=\ov{\Omega} \label{O6projSU3} \ ,\\
O_7 :& \quad \sigma(J)=J,\ \sigma(\Omega)=-\Omega \label{O7proj} \ .
\eea
Note that $A$ is always assumed even under $\sigma$.

\subsection*{Structure group compatibility conditions}

Forms defining an SU(3) structure have to satisfy the following compatibility conditions
\beq
J\w \Omega=0 \ ,\ \i \Omega \w \ov{\Omega}= \frac{4}{3} J^3 \neq 0 \ ,\label{compatSU3}
\eeq
while those for an SU(2) structure are given by
\bea
& j^2=\frac{1}{2} \omega\w \ov{\omega} \neq 0 \ ,\ j\w \omega=0\ ,\ \omega\w \omega=0 \label{compatjo}\\
& z\vee \omega= 0 \ ,\ z\vee j=0 \label{compatz} \ .
\eea
In our vacua, we will consider a parametrization of the structure forms in terms of a basis of one-forms, (1,0) or (0,1) with respect to the almost complex structure, as discussed around \eqref{SU3forms} and \eqref{SU2forms}. The compatibility conditions will then be automatically satisfied, and the corresponding metric $g_{ab}$ will be easily built.

\subsection*{SUSY conditions and Hodge star}

The SUSY conditions \eqref{SUSY} can be developed according to the choice of pure spinors \eqref{purespinorstruct} and the theory. We start with the case of an orthogonal SU(2) structure, that will only be needed in type IIA. We fix the phase $\theta_-=\pi$: this phase is unphysical \cite{Koerber:2007hd} and can always be absorbed in a redefinition of $z$, as can be seen in the pure spinors. It was taken to be $\frac{\pi}{2}$ in \cite{Grana:2006kf}. After a few manipulations, the SUSY conditions become
\bea
& \d (e^{3A-\p} \omega)=0 \label{SUSYSU2}\\
& \omega\w \d (\frac{1}{2} z\w \ov{z})= H\w \omega \nn\\
& \d (e^{2A-\p} \re(z))=0 \nn\\
& \d (e^{2A-\p} \im(z)\w j)= e^{2A-\p} H\w \re(z) \nn\\
& \re(z)\w \d (\frac{1}{2} j^2)= H\w \im(z)\w j \nn\\
& F_6=0 \nn\\
& \d (e^{4A-\p} \im(z))= e^{4A} * F_4 \nn\\
& \re(z)\w \d (e^{2A} j)= e^{2A} H\w \im(z) - e^{2A+\p} * F_2 \nn\\
& \im(z)\w j\w \d (\frac{e^{2A}}{2} j)= -\frac{1}{2} e^{2A} H\w \re(z) \w j + e^{2A+\p} * F_0 \ . \nn
\eea
We now turn to an SU(3) structure. The phases are fixed as discussed above for the orientifold projection. In the case of an $O_5$, the SUSY conditions are then equivalent to
\bea
& e^{\phi}=g_s e^{2A} \label{SUSYSU3O5}\\
& H=0,\ F_1=0,\ F_5=0 \nn\\
& \d (e^A \Omega)=0 \nn\\
& \d (J^2)=0 \nn\\
& \d(e^{2A} J)= -g_s e^{4A} * F_3 \ . \nn
\eea
For an $O_6$, they are equivalent to
\bea
& e^{\phi}=g_s e^{3A} \label{SUSYSU3O6}\\
& H=0,\ F_0=0,\ F_4=0,\ F_6=0 \nn\\
& \d (J)=0 \nn\\
& \d (e^{-A} \im( \Omega))=0 \nn\\
& \d(e^{A}  \re( \Omega))= -g_s e^{4A} * F_2 \ . \nn
\eea
For an $O_7$, we choose $e^{\phi}=g_s e^{4A}$. The SUSY conditions are then equivalent to
\bea
& \d (e^{-A} \Omega)=0 \label{SUSYSU3O7}\\
& \d(e^{-2A} J)=0 \nn\\
& H\w \Omega=0,\ H\w J=0 \nn\\
& H= -g_s e^{4A} * F_3, \ F_5=0 \nn\\
& \frac{1}{2}\d(J^2)= g_s e^{4A} * F_1 \ . \nn
\eea

Obtaining the RR fluxes from the above requires the six-dimensional Hodge star: we formulate it here in the $e^a$ basis, with the diagonal $g_{ab}$ \eqref{defmetric}
\beq
*(e^{a_1} \w \dots \w e^{a_k})= \frac{\epsilon_{a_1 \dots a_6}}{\sqrt{|g|}} g_{a_{k+1} a_{k+1}} \dots g_{a_{6} a_6} e^{a_{k+1}} \w \dots \w a^{a_6} \ , \label{Hodge}
\eeq
which is understood without summation in the r.h.s. but with fixed indices ($\epsilon$ is here only a sign, $\epsilon_{1 \dots 6}=1$), and $|g|$ is the absolute value of the determinant of the metric. We consider an Euclidian space signature, giving for a $p$-form: $ **A_p= (-1)^{p(6-p)} A_p$.

\subsection*{RR Bianchi identities}

To fix the notations of the r.h.s. of the BI \eqref{BI}, one should derive it from the sources action with appropriate conventions. We take here a more pragmatic approach following \cite{Grana:2006kf, Andriot:2010sya}. We use the convention \eqref{defvolbot} for ${\rm vol}_{\bot}$, which ensures that we read-off the right signs of the charges in the BI, provided some conditions are satisfied; this formulation is enough for our purposes.

In our vacua, we only face cases with sources $O_p$ and $D_p$ of fixed $p$ along one ${\rm vol}_{||}$, sourcing one flux $F_k$, and with most of the time $H=0$. Given the sign $\lambda(F_k)= (-1)^{[\frac{k}{2}]} F_k$ where $[.]$ denotes the integer part, we then take as a convention
\beq
(-1)^{[\frac{k}{2}]} ~ {\rm vol}_{\bot} \w {\rm vol}_{||} = {\rm vol}_6 \ . \label{defvolbot}
\eeq
In the cases considered, it allows the following equalities
\bea
\int_{\mmm} \frac{e^{4A}}{8} |F_k|^2\ {\rm vol}_6 &= \int_{\mmm} \frac{e^{4A}}{8} F_k \w * F_k \\
&= \int_{\mmm} (-1)^k (-1)^{[\frac{k}{2}]} F_k \w \d (e^{3A-\phi} \im (\Phi_2)|_{6-k-1}) \nn\\
&= - \int_{\mmm} e^{3A-\phi} (-1)^{[\frac{k}{2}]} \d F_k \w \im (\Phi_2)|_{6-k-1} \nn\\
&= - \int_{\mmm} e^{3A-\phi} c_p \left( - \sum_{O} 2^{p-5}~ \delta(y) + \sum_{D} \delta(y)\right) \ (-1)^{[\frac{k}{2}]} ~ {\rm vol}_{\bot}  \w \im (\Phi_2)|_{6-k-1} \nn\\
&= - \int_{\mmm} e^{3A-\phi} c_p \left( - \sum_{O} 2^{p-5}~ \delta(y) + \sum_{D} \delta(y)\right) \ (-1)^{[\frac{k}{2}]} ~ {\rm vol}_{\bot}  \w P\left[\im (\Phi_2)\right] \nn\\
&= - \int_{\mmm} \frac{e^{4A-\phi}}{8} c_p \left( - \sum_{O} 2^{p-5}~ \delta(y) + \sum_{D} \delta(y)\right) \ (-1)^{[\frac{k}{2}]} ~ {\rm vol}_{\bot}  \w {\rm vol}_{||} \nn\\
&= - \int_{\mmm} \frac{e^{4A-\phi}}{8} c_p \left( - \sum_{O} 2^{p-5}~ \delta(y) + \sum_{D} \delta(y)\right) \ {\rm vol}_6 \nn
\eea
where we also used the Hodge star \eqref{Hodge}, the SUSY condition (the sign $(-1)^k$ is the theory dependent one), the fact that $\mmm$ has no boundary, the BI \eqref{BI} and the calibration \eqref{calib}. In the smeared limit ($\delta \rightarrow 1, A\rightarrow 0$), this convention provides precisely the right signs: the first integrand and $c_p$ being positive, one gets that the number $( - 2^{p-5}~ N_O + N_D)$ should be negative as usually required (the need for orientifolds). The signs are preserved in the non-smeared case. While providing the right signs for the charges, the convention \eqref{defvolbot} fixes as well the sign ambiguity when ordering the subvolumes. For an extension of this reasoning to a more general approach introducing currents localising the sources, see \cite{Grana:2006kf} or appendix B.3 of \cite{Andriot:2010sya} and references therein.

Having fixed the signs and notations, the BI boils down to a differential equation that is in practice never solved: the warp factor is always considered to be the solution. The difficulty in the BI is rather to get no other terms than the ones transverse to the desired sources. When looking for vacua, getting rid of undesired terms often imposes further constraints.

\section{T-duality along a single direction}\label{ap:Td}

We motivated in section \ref{sec:Td} the importance to study the effect of T-duality on the vacua obtained in section \ref{sec:newsol} on solvmanifolds corresponding to $s_3$. We gave there arguments and necessary material to tackle the T-duality along both directions of a pair with equal radii. We now build on this to look at the T-duality along a single direction in such a pair, say $\d y^1$. In the smeared limit, i.e. without warp factor, which is enough for our purposes, no field depends on $y^1$ so it can be viewed as an isometry direction, allowing the T-duality.\footnote{When the solvmanifold is not a torus, $\d y^1$ is not globally well-defined, so T-dualising along such a direction might be questionable. A similar situation occurs though when T-dualising the Heisenberg manifold back to a torus. The questions raised in this appendix remain of interest, so we pursue this study. Note that there is no such doubt when T-dualising along both directions of the pair since $\d y^1 \w \d y^2=e^1 \w e^2$ is globally defined.} From the perspective of the metric \eqref{funnymetric}, the result of the T-duality is then simple: one $\tilde{g}_{11}$ gets inverted. The T-dual space is however not clearly identified: on the one hand, we recall the discussion of section \ref{sec:Td} on the (non)-relation to the torus, but on the other hand, the one-forms of the solvmanifold can this time not be rebuilt easily in the metric. So it is more instructive to take the (generalized) vielbeins point of view. First, note that the metric determinant is never vanishing here; after the T-duality it is in particular equal to one. So the same should hold for the vielbein $e$ and T-dual $e'$: this is a minimal requirement to have consistent vielbeins. When computing $\eee O$ and reading-off the upper left block, one gets however a determinant zero matrix. This typical situation is cured by acting on the left with a $K$, as described in section \ref{sec:Td}. A first standard choice is $K=O$, i.e. we look at $O \eee O$: the resulting upper left block is given by
\beq
\cos (q y^5) \begin{pmatrix} \frac{1}{R} & 0 \\ 0 & R \end{pmatrix} \ .\label{vielbeinT}
\eeq
Its determinant indicates however once more that \eqref{vielbeinT} is not a suitable T-dual vielbein $e'$; one needs a further $K$. To that end, the appendix B of \cite{Andriot:2010ju} is helpful, since similar computations were made there.\footnote{In that appendix were also studied T-duals of $s_3$ vacua. We believe however that some of the results there are not correct: our main criticism is that the vielbein there takes $q=1$, which makes the solvmanifold a torus, falsifying the interpretations. The globally definedness of $\cos (q y^5)$ is used to study $K$, e.g. in the last sentence of that appendix, but we now know that away from the torus, this function is not globally defined. In addition, the T-dual space is studied around (B.3) and (B.4), in a case where the two radii are not equal ($t_1 (\tau_2^1)^2 \neq t_2$). A single T-duality is argued to lead to a manifold corresponding to $s_1$, with a metric involving a function $G$ that depends on $\cos (q y^5)$. Again, if that function is globally defined, the T-duality essentially relates a torus to another torus; if it is not globally defined, $G$ and the metric are not either, so the argument given there does not hold. We thus start over the analysis here.} At this stage, let us discuss the other blocks of the generalized vielbein. The lower left one should depend on a $b$-field (as in standard supergravity), while the upper right one can correspond to a bivector $\beta$ on which $\beta$-supergravity was built \cite{Andriot:2011uh, Andriot:2013xca}.\footnote{The off-diagonal blocks going with \eqref{vielbeinT} are not antisymmetric, except for the special case $R=1$. In general, these blocks can then not be interpreted directly as $b$ or $\beta$.} Having this last block non-zero is commonly related to having non-geometric fluxes or understood as a sign of non-geometry. It is argued in \cite{Grana:2008yw} that if a $K$ sets this upper right block to zero and is not globally defined, then one is not facing a geometric background, but maybe a non-geometric one; we will use this argument. Upon some restriction, the generalized vielbein without a $b$-field but with a $\beta$ was shown in \cite{Andriot:2014uda} to provide a geometric description of a non-geometric background; reaching such a generalized vielbein takes the form of a transformation by $K$, as shown e.g. in (C.9) of \cite{Andriot:2013xca}. Here, by acting with $K$, we want the upper left block of the generalized vielbein to provide a suitable, consistent, vielbein $e'$; the other blocks may contain $b$ or $\beta$ (or both as in \cite{Aldazabal:2011nj, Geissbuhler:2013uka}) that could then be studied and related to (non)-geometric fluxes. To do so, a first attempt is to reach the following vielbein, that would correspond to a torus with some radii
\beq
\begin{pmatrix} \frac{1}{R} & 0 \\ 0 & R \end{pmatrix} \ .
\eeq
This is actually achieved with an O(2)$\times$O(2) $K$, which however depends on $\cos (q y^5)$ and is thus not globally defined. In addition, the resulting off-diagonal blocks of the generalized vielbein are set to zero by this $K$. Following the argument of \cite{Grana:2008yw}, the T-dual background is then not geometric; whether it is non-geometric remains to be studied, following for instance the approach of section 4.2.2 of \cite{Andriot:2014uda} and looking at how transition functions get transformed under the T-duality. Studying $s_1$ in \cite{Hassler:2014sba}, it was argued from a four-dimensional T-duality that the T-dual should be a torus with an $H$-flux and a non-geometric $Q$-flux; here, the off-diagonal blocks being zero, we do not manage to see any flux.\footnote{Note that a four-dimensional T-duality is typically along flat indices, meaning here along e.g. $e^1$ instead of $\d y^1$. But it is not clear how to perform such a transformation at the world-sheet level.} A second attempt worth being mentioned is to reach the following vielbein, corresponding to the initial solvmanifold but with different radii
\beq
\begin{pmatrix} \frac{1}{R} & 0 \\ 0 & R \end{pmatrix} \begin{pmatrix} \cos (q y^5) & -\sin (q y^5) \\ \sin (q y^5) & \cos (q y^5) \end{pmatrix} \ .
\eeq
This is achieved by a transformation on the left, which is however not an O(2)$\times$O(2) $K$. We refrain from looking further for a more satisfying basis for the T-dual generalized vielbein. To conclude, we can say at least that a T-duality along a single direction of the pair does not lead to a geometric vacuum on a torus.

\end{appendix}

\newpage

\providecommand{\href}[2]{#2}\begingroup\raggedright

\endgroup

\end{document}